\documentclass[pdflatex,sn-basic]{sn-jnl}

\usepackage{graphicx}
\usepackage{subfigure}
\usepackage{ amssymb}
\usepackage { amsmath }

\usepackage { algorithm }
\usepackage { algpseudocode }

\usepackage {adjustbox}
\usepackage {multirow}

\usepackage[figuresright]{rotating}
\usepackage{booktabs}

\usepackage{natbib}
\usepackage{hyperref}

\usepackage{lineno}
\usepackage{supertabular}
\usepackage{ulem}
\usepackage{setspace}
\usepackage[misc]{ifsym}

\jyear{2023}%

\theoremstyle{thmstyleone}%
\theoremstyle{thmstyletwo}%

\theoremstyle{thmstylethree}%

\begin{document}
\title[Attention-based Graph Neural Networks: A Survey]{Attention-based Graph Neural Networks: A Survey}


\author[1]{Chengcheng Sun\footnote{Chengcheng Sun and  Chenhao Li have contributed equally to this work.}}

\author[1]{Chenhao Li}

\author[1]{Xiang Lin}

\author[1]{Tianji Zheng}

\author[1,2]{Fanrong Meng}

\author[1]{Xiaobin Rui}

\author*[1,2]{Zhixiao Wang}\email{zhxwang@cumt.edu.cn}

\affil[1]{\orgdiv{School of Computer Science and Technology}, \orgname{China University of Mining and Technology}, \orgaddress{\street{Xuzhou}, \city{Jiangsu}, \postcode{221116}, \country{China}}}

\affil[2]{\orgdiv{Mine Digitization Engineering Research Center}, \orgname{Ministry of Education of the People's Republic of China}, \orgaddress{Xuzhou, Jiangsu, \postcode{221116}, \country{China}}}



\abstract{
Graph neural networks (GNNs) aim to learn well-trained representations in a lower-dimension space for downstream tasks while preserving the topological structures. In recent years, attention mechanism, which is brilliant in the fields of natural language processing (NLP) and computer vision (CV), is introduced to GNNs to adaptively select the discriminative features and automatically filter the noisy information. To the best of our knowledge, due to the fast-paced advances in this domain, a systematic overview of attention-based GNNs is still missing. To fill this gap, this paper aims to provide a comprehensive survey on recent advances in attention-based GNNs. Firstly, we propose a novel two-level taxonomy for attention-based GNNs from the perspective of development history and architectural perspectives. Specifically, the upper level reveals the three developmental stages of attention-based GNNs, including graph recurrent attention networks, graph attention networks, and graph transformers. The lower level focuses on various typical architectures of each stage. Secondly, we review these attention-based methods following the proposed taxonomy in detail and summarize the advantages and disadvantages of various models. A model characteristics table is also provided for a more comprehensive comparison. Thirdly, we share our thoughts on some open issues and future directions of attention-based GNNs. We hope this survey will provide researchers with an up-to-date reference regarding applications of attention-based GNNs. In addition, to cope with the rapid development in this field, we intend to share the relevant latest papers as an open resource at  \href{https://github.com/sunxiaobei/awesome-attention-based-gnns}{https://github.com/sunxiaobei/awesome-attention-based-gnns}.
}



\maketitle

\section{Introduction}
\label{intro}

As a typical non-Euclidean data structure, graphs are ubiquitous in the real world, such as social networks, transportation networks, citation networks, and protein interaction networks. 
Recent deep learning research on graphs has attracted more and more attention due to the rich expressive power of deep learning models \citep{1_graph-survey_2020}. 
Graph neural networks (GNNs \citep{63_GNN-model_2008,14_spectral_2014}), as a kind of deep graph representation learning method, aim to learn nodes/edges/graphs-level representations in a lower-dimension space for various downstream tasks (e.g., node classification \citep{2_GCN_2017}, link prediction \citep{3_Simple-HGN_2021}, community detection \citep{4_comprehensive-survey_2022}, and graph classification \citep{5_Graphormer_2021}). 
Different from the traditional graph embedding methods (i.e., matrix factorization \citep{6_linear_2000} and random walk-based methods \citep{7_deepwalk_2014,8_line_2015,9_node2vec_2016}), GNNs learn both the structural and attribute features of a graph through different neural networks. 
Due to this inherent advantage, GNNs can naturally be used to deal with graph-structured data with a large amount of attribute information \citep{10_effective_2021}.

Existing GNNs \citep{11_GNN-survey_2020}, originating from graph convolution \citep{2_GCN_2017}, graph diffusion \citep{12_diffusion_2019}, graph attention \citep{13_GAT_2018} or other various mechanisms, are usually divided into two classical categories: spectral domain \citep{14_spectral_2014,15_convolutional_2016,2_GCN_2017} and spatial domain \citep{16_inductive_2017,13_GAT_2018,17_how-powerful_2018}. 
Spectral graph neural networks realize topological convolution operations based on spectral graph theory and graph signal processing. Typical methods include SpectralCNN \citep{14_spectral_2014}, ChebNet \citep{15_convolutional_2016}, and GCN \citep{2_GCN_2017}. Spectral GNN is based on the Fourier transform and strongly depends on the graph Laplacian matrix \citep{2_GCN_2017}. 
Once the graph structure is determined, it is difficult to extend such methods to another new graph. 
Spatial graph neural networks directly aggregate and update the information of nodes from neighborhoods iteratively under message-passing neural networks (MPNN \citep{18_neural-message-passing_2017}). Considering the attributes of nodes/edges/graphs, Graph Networks (GraphNet \citep{2018_GraphNet}) generalizes and extends various graph neural networks under message-passing mechanism and supports constructing complex architectures from simple building blocks. 
Most GNNs treat neighborhood nodes equally without considering the contributions of different nodes when aggregating and propagating information. 
However, real-world graphs are often noisy with connections between unrelated nodes, which causes GNNs to learn suboptimal representations without distinguishing nodes in neighborhoods \citep{19_SuperGAT_2021}.

To avoid noisy signals, GAT \citep{13_GAT_2018}, a classical spatial graph neural network, introduces an attention mechanism into the graph neural network to learn the importance of different neighbors. 
Thereafter, attention becomes a popular mechanism used in a wide range of graph deep neural network architectures for better representation \citep{19_SuperGAT_2021}. 
Till now, there are a large number of attention-based graph neural networks. Attention in GNNs can be applied to different node neighborhoods, different parts of structures, and different representations of nodes, edges, and graphs.

In 2019, a survey reviews several attention models in graphs from the perspective of graph types, attention types, and task types \citep{20_attention-survey_2019}. After that, reviews of attention models in graphs have been vacant. Most reviews focus more on the taxonomy and application of existing GNNs \citep{11_GNN-survey_2020,2022_GNN_survey_zhou,2021_GNN_survey_Geo,169_GNN-nlp-survey_2021,2022_GNN4RS_survey_Gao,171_GNN-traffic_2021,2022_GNN4IoT_survey_Cini}, and they also describe GNNs in both the spectral domain and spatial domain. A comprehensive survey \citep{11_GNN-survey_2020} divides the state-of-the-art GNNs into four categories, including recurrent graph neural networks, convolutional graph neural networks, graph autoencoders, and spatial-temporal graph neural networks. To better address the challenges that exist in GNNs, a survey discusses several fundamental challenge problems from both theoretical and practical perspectives \citep{2021_GNN_survey_Geo}. Other reviews cover more focused aspects of specific application domains, such as GNN4NLP \citep{169_GNN-nlp-survey_2021}, GNN4RS \citep{2022_GNN4RS_survey_Gao}, GNN4Traffic \citep{171_GNN-traffic_2021}, and GNN4IoT \citep{2022_GNN4IoT_survey_Cini}. Recently, an overview provides a review of various transformers for graphs from an architectural perspective \citep{2022_Graph_Transformer_Min}. Several attention-based GNNs are usually covered in the above reviews on GNNs, but they only list a few of the core literature on attention-based GNNs without further classification. By contrast, we aim to provide a systematic overview of attention-based GNNs from development history and architectural perspectives in this survey. Considering the rapid evolution of this field, it’s imperative to sort out the existing attention-based GNNs and systematically analyze advanced methods for both academics and practitioners. However, to the best of our knowledge, an up-to-date, and comprehensive overview of attention-based GNNs is missing. To fill this gap, we present a novel two-level taxonomy from both developmental stages and architectural perspectives to systematically review attention-based GNNs.

\begin{figure}[!htbp]
    \centering
    \includegraphics[width=1\linewidth]{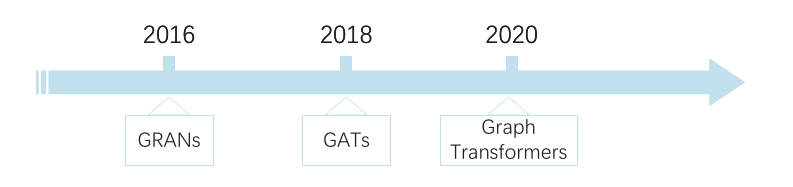}
    \caption{Three developmental stages of attention-based graph neural networks}
    \label{fig:stage_attention_gnn}
\end{figure}

Existing attention-based graph neural networks can be divided into three stages, including Graph Recurrent Attention Networks (GRANs \citep{21_GGNN_2016,22_GaAN_2018,23_GRAN_2019}), Graph Attention Networks (GATs \citep{13_GAT_2018,19_SuperGAT_2021}), and Graph Transformers \citep{25_UGformer_2019,24_Graph-bert_2020}, as shown in Figure \ref{fig:stage_attention_gnn}. 
GRANs introduce the attention mechanism and RNN into graph neural networks. Due to the inherent limitations of RNN, the calculation of each step of GRANs usually depends on the results of the previous step. On the contrary, GATs do not need to rely on the previous results and can implement parallel operations. Most GATs focus on nodes in local neighborhoods with local attention, which distinguish the importance of different neighbors. However, GRANs and GATs also suffer from a common limitation: they cannot capture remote dependencies. Consider capturing remote messages, Graph Transformers can learn higher-order graph information directly with global attention. Each stage has its representative and typical methods. Therefore, we further categorize these methods from an architectural perspective for each stage.

In GRANs, most methods are inspired by attention-based Recurrent Neural Networks (RNNs). Therefore, we naturally divide these methods into two subclasses. 
One is based on Gated Recurrent Units (GRU) \citep{22_GaAN_2018}, and the other is attention-based Long Short-Term Memory (LSTM) \citep{26_JK-Net_2018}. However, both of those methods are suffered from the long-time dependency bottleneck and the order problem that exists in RNNs.

GATs assign different weights to nodes in the feature aggregation steps according to their influences. GAT \citep{13_GAT_2018} is the pioneer of this stage. Later, a great variety of GATs, including C-GAT \citep{27_C-GAT_2019}, GATv2 \citep{28_GATv2_2021}, and SuperGAT \citep{19_SuperGAT_2021}, adopt different attention strategies to GNNs. 
We distinguish GATs from the perspective of whether the attention mechanism is in the neural network layer, namely intra-layer GATs and inter-layer GATs. 
For intra-layer GATs, the attention function is used to calculate the weights of different nodes in local neighborhoods, and then dynamically update the representation of nodes. 
For inter-layer GATs, attention is usually regarded as an operation of feature combination, which is used to select features from different levels, different channels, different views, or different time slices.

Graph Transformers \citep{156_GROVER_2020} can learn higher-order graph properties directly, different from previous methods with local attention. In the past two years, Graph Transformers have developed rapidly in the field of graph deep learning, especially in the task of graph classification. However, the globally-connected self-attention mechanism in graph transformers makes it necessary to update the weights of the whole network continuously in the process of end-to-end model training \citep{5_Graphormer_2021,30_transformers-generalize_2021}. 
This prevents them from exploiting sparsity in the graph topology, leading to excessively high computational complexity.

In a word, this paper aims to provide a systematic and comprehensive review of contemporary attention-based graph neural networks. Our contributions can be summarized as follows.
\begin{itemize}
\item This paper proposes a novel two-level taxonomy for attention-based GNNs from the perspective of development history and architecture. Specifically, the upper level reveals the three developmental stages of attention-based GNNs, including GRANs, GATs, and Graph Transformers. The lower level focuses on various typical architectures of each stage.

\item This paper comprehensively and systematically summarizes the latest works on attention-based GNNs, which makes up for the lack of literature in this hot direction. For each sub-category, we provide a detailed introduction and an in-depth comparison to reveal the advantages and disadvantages of various models.

\item This paper also provides open issues and challenges of attention-based GNNs as insights for future research directions to advance this field, which will provide researchers with an up-to-date reference about attention-based GNNs.
\end{itemize}

The rest of this survey paper is organized as follows. Section \ref{Preliminaries_Notations} gives the preliminaries and notations of attention-based GNNs. Section \ref{Taxonomy} provides a two-level taxonomy from the history and architectural perspectives. Section \ref{GRANs} presents a technical overview of GRANs. We introduce the GATs in both Section \ref{Intra-Layer GATs} and Section \ref{Inter-Layer GATs}. Section \ref{Graph Transformers} overviews 
Graph Transformers. In Section \ref{Comparison}, we summarize and compare the  characteristics of different models in subclasses. We suggest promising future research directions in Section \ref{Open Issues and Future Directions} and Section \ref{Conclusion} comes to the conclusion of this paper.

\section{Preliminaries and Notations}
\label{Preliminaries_Notations}

\subsection{Graph}
\begin{figure}[!htbp]
    \centering
    \includegraphics[width=0.8\linewidth]{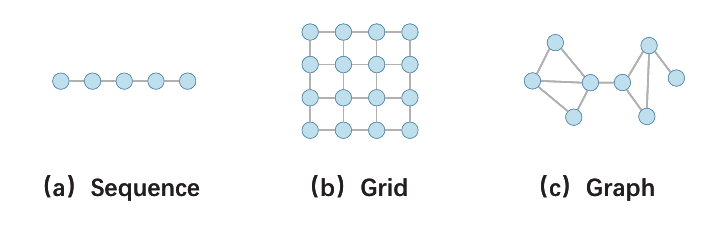}
    \caption{Examples of sequence-structured, grid-structured, and graph-structured data}
    \label{fig:data_example}
\end{figure}

A graph is a data structure defining a set of nodes and their relationships. As shown in Figure \ref{fig:data_example}, unlike sequence-structured data in NLP and grid-structured data in CV, graph-structured data is extremely complex. Especially, the number of neighbor nodes for each node is irregular. With the growth of the graph scale, the number of nodes in the graph often increases exponentially.

Let $G = \langle V, E \rangle$ represents an attributed graph with a feature matrix as $X$ where $V=\left\{v_1, v_2,\cdots, v_n\right\}$ refers to the set of nodes, $E = \left\{e_1, e_2,\cdots, e_m\right\}$ refers to the set of edges between nodes.
We denote $n=\vert V \vert$ as the number of nodes and $ m=\vert E \vert$ as the number of edges. $X\in\mathbb{R}^{n\times dim}$ represents the feature matrix and its row element $\mathbf{x}_i\in\mathbb{R}^{dim}$ represents the feature vector of node $v_i$ with the dimension $dim$.  
The adjacency matrix could represent the connection relationship between nodes, defined as $A\in\left\{0,1\right\}^{n\times n}$, where $A_{ij}=1$ means that there is an edge $e_{ij}$ between node $v_i$ and $v_j$. We use $d_i=\sum_{j\in\mathrm{\Gamma}_i} A_{ij}$ to denote the degree of node $v_i$ , where $\mathrm{\Gamma}_i$ is the neighborhood nodes of node $v_i$. 
For a multi-relational graph, we extend the edge notation with edge type $r\in R$, as $e_{ij}^r$. Correspondingly, edge and graph can have attributes as well, defined as $X_e$, $X_g$. If not specified, for convenience we default $X$ to represent the node attribute matrix.

\subsection{Graph Neural Networks}

\begin{figure}[!htbp]
    \centering
    \includegraphics[width=1\linewidth]{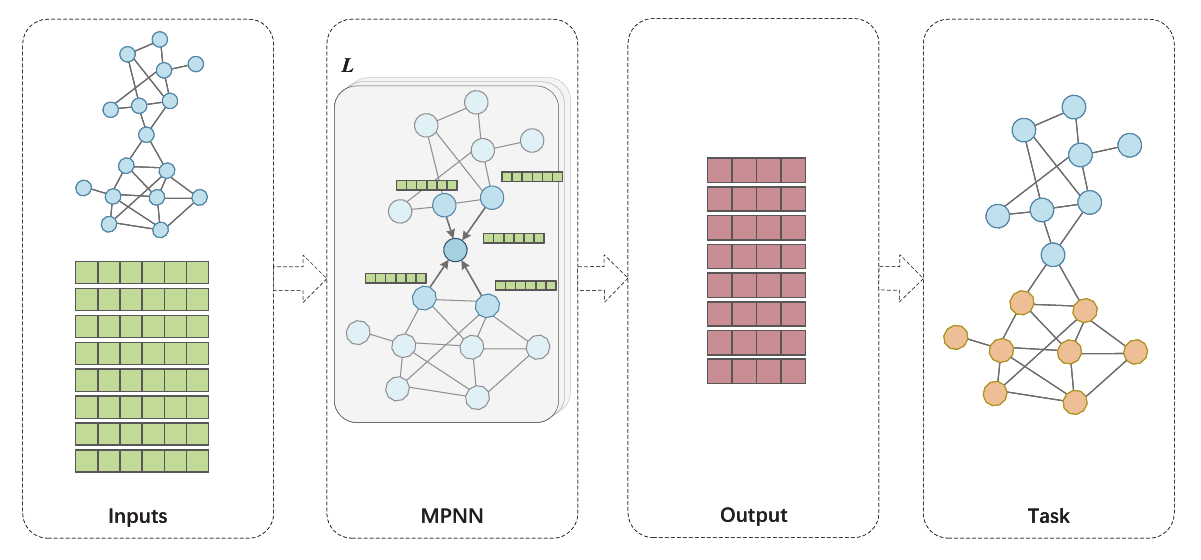}
    \caption{The architecture of GNNs with the message-passing mechanism}
    \label{fig:MPNN}
\end{figure}

Graph Neural Networks, a series of methods in Graph Representation Learning (GRL) based on deep learning, take a source graph $G$ as input, with an adjacency matrix $A$ and feature matrix $X$. GNNs aim to learn a potential representation embedding vectors $Z\in\mathbb{R}^{n\times{dim}^\prime}$ in low dimensionality, i.e., $dim^\prime\ll dim$, which will play a key role in downstream tasks, such as node classification, link prediction, and graph classification. See Figure \ref{fig:MPNN} for an illustration.

From a spatial perspective, MPNN \citep{18_neural-message-passing_2017} can be seen as a more constrained instance of GraphNet \citep{2018_GraphNet}, considering only the attributes of the nodes. They propagate information over the graph by a local diffusion process \citep{2020_GraphNet_Spectral}. The GraphNet block contains three update functions and three aggregation functions, while MPNN has only one pair of such functions.

Taking node features $X$ and graph structure $A$ as inputs, GNNs can be defined as $Z^{out}=\mathcal{F}\left(X,\ \ A,\ \ \Theta\right)$, where $\Theta$ refers to the learnable weight parameter and $\mathcal{F}\left(\cdot\right)$ denotes the GNN encoders such as GCNs, GATs even MLPs. Mathematically, we can define the general framework of graph neural networks with the message-passing mechanism as follows:

\begin{equation}
    M_{v_i}^l=\ {Agg}^{l}\left(\left\{H_{v_j}^{l-1},\forall v_j\in\mathrm{\Gamma}_i\right\}\right)
\label{eq:eq_mpnn_agg}
\end{equation}
\begin{equation}
    H_{v_i}^l=\ {Comb}^l\left(H_{v_i}^{l-1},\ {\ M}_{v_i}^l\right)
\label{eq:eq_mpnn_comb}
\end{equation}
where $H^l{\in\mathbb{R}}^{n\times dim}$ represents the node features in the $l^{th}$ layer starting from the initial node features $H^0=X$. $\mathrm{\Gamma}_i$ denotes the neighborhood node $v_i$ with a set of neighbors. 
$M_{v_i}^l$ refers to the received messages from neighbors in the neighborhood $\mathrm{\Gamma}_i$ of the current node $v_i$ in the $l^{th}$ layer. From the view of the message-passing mechanism, GNN can be defined as two important functions: $Agg\left(\cdot\right)$ and $Comb\left(\cdot\right)$.
$Agg\left(\cdot\right)$ is the aggregator function to aggregate the messages from the neighbors of each node in the graph, while $Comb\left(\cdot\right)$ is the combined (or update) function to update the node representations by combining the received messages from neighbors and the representations of the current node in the former layer.

For the node classification task in Figure \ref{fig:MPNN}, we can take $Z^{out}=H^L$ as the output of graph neural networks after $L$ layers. For graph-level tasks, an additional readout function $Read\left(\cdot\right)$ should be defined to read out the final representation $H_G$ of the entire graph or subgraph. There are a series of readout functions, such as average, sum, and max functions, which can be defined as below:

\begin{equation}
    H_G=Read(\{H_v^l,\ \forall\ v\in V\})
\label{eq:eq_read}
\end{equation}

Take GCN for example, which is the most popular graph neural network due to its ease of understanding from the spatial domain as well as the theoretical basis from the spectral domain \citep{31_survey-graph-learning-2021}. 
GCN iteratively aggregates and updates the representations of nodes in the graph through a propagation rule \citep{2_GCN_2017} defined as:

\begin{equation}
    H^l=\sigma\left({\widetilde{D}}^{-\frac{1}{2}}\widetilde{A}{\widetilde{D}}^{-\frac{1}{2}}H^{l-1}\Theta^l\right)
\label{eq:eq_gcn_propagation}
\end{equation}
where $\widetilde{A}=A+I_n$ denotes the adjacency matrix with self-loop and $\widetilde{D}$ represents the corresponding diagonal degree matrix. 
$I_n$ refers to the identity matrix, and ${\widetilde{D}}^{-\frac{1}{2}}\widetilde{A}{\widetilde{D}}^{-\frac{1}{2}}$ is a renormalization trick inspired by the symmetric normalized graph Laplacian $L^{sn}=I_n-D^{-\frac{1}{2}}AD^{-\frac{1}{2}}$. 
$\Theta$ is a trainable weight parameter and $\sigma\left(\cdot\right)$ is a non-linear activation function $ReLU\left(x\right)=\max(0,x)$. 
For the message-passing mechanism in GCN, we rewrite the above formula from the view of nodes as follows:

\begin{equation}
    H_{v_i}^l=\sigma\left(\sum_{j\in\mathrm{\Gamma}_i}\frac{{\widetilde{A}}_{ij}}{\sqrt{d_id_j}}{H_{v_j}^{l-1}\Theta}^l+\ \frac{{\widetilde{A}}_{ii}}{\sqrt{d_id_i}}{H_{v_i}^{l-1}\Theta}^l\right) 
\label{eq:eq_gcn_mp}
\end{equation}
where $d_i$ refers to the degree of node $v_i$ . $Agg$ function in GCN is defined as the average of the neighbor node representations with a normalization constant based on degree. $Comb$ function is defined as a simple summation function.

\subsection{Attention Mechanisms}

Attention is a mechanism based on the recognition process of the human visual system, which imitates the human cognitive awareness about specific information to focus more on the critical aspects of data \citep{32_survey-CV_2022,33_deep-learning-survey_2021}. 
The attention mechanisms have been successfully applied to various application fields in CV, NLP, and GRL. 
Inspired by attention in CV \citep{32_survey-CV_2022} and NLP \citep{34_attention-models-survey_2021}, we give a generalized definition of attention and then decompose the attention process into three effective functions, i.e., alignment function, distribution function, and weighted sum function.

The attention mechanism with different attention functions can be defined under a generalized framework:

\begin{equation}
    Attention=f\left(g\left(X\right),\ X\right)
\label{eq:eq_attention}
\end{equation}
where $g\left(\cdot\right)$ refers to an attention function to generate attention which corresponds to capturing the important regions in the visual scene \citep{32_survey-CV_2022}. $f\left(\cdot\right)$ means processing input data $X$ to obtain important information based on the attention function $g\left(\cdot\right)$. 

\begin{figure}[!htb]
    \centering
    \includegraphics[width=1\linewidth]{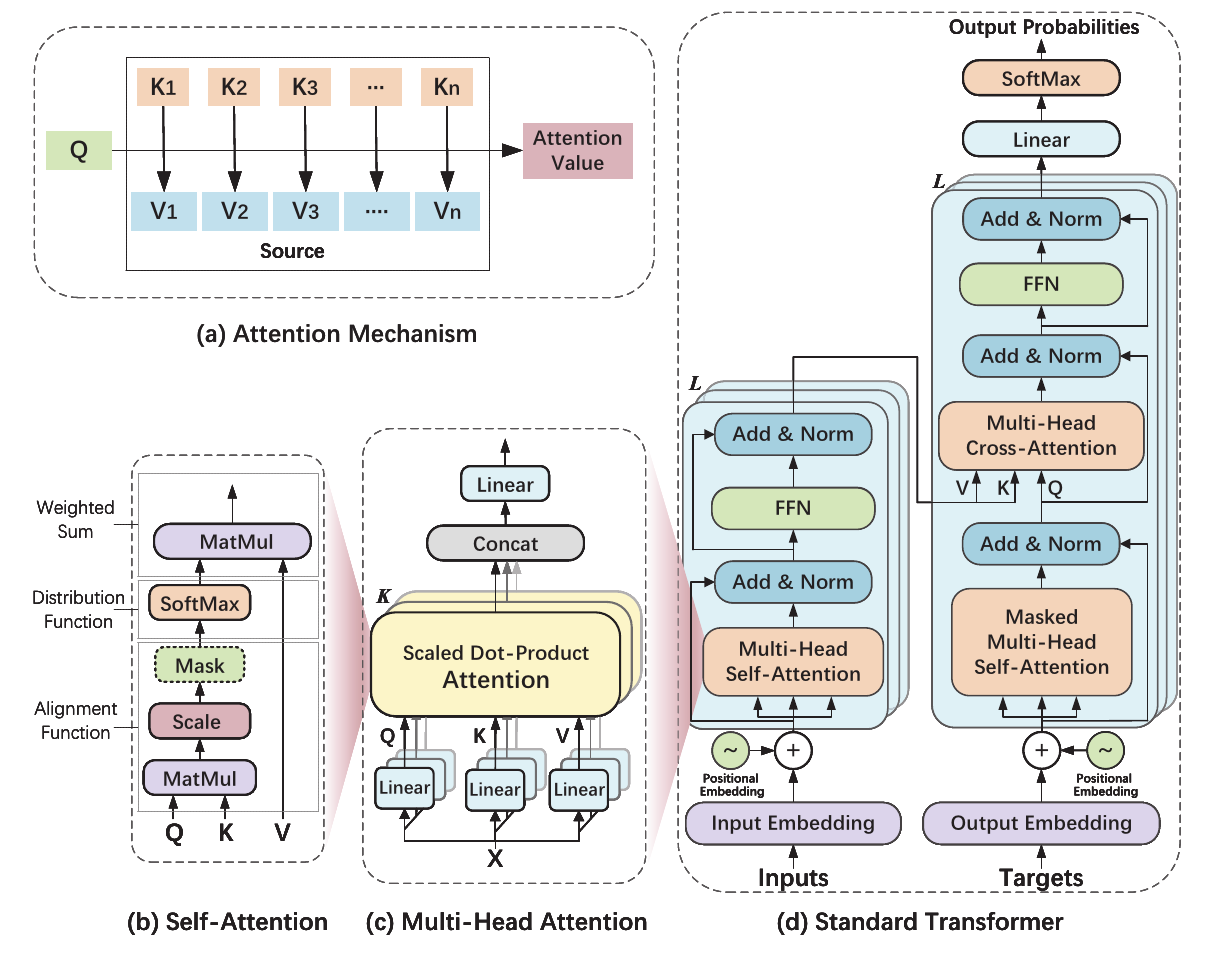}
    \caption{The architecture of attention, self-attention, multi-head attention, and standard transformer}
    \label{fig:attention_architecture}
\end{figure}

In more detail, the attention mechanism (shown in Figure \ref{fig:attention_architecture}(a)) can be seen as a mapping of a sequence of keys ${K}$ to an attention distribution $\alpha$ according to queries ${Q}$, and the keys have one-to-one corresponding values ${V}$ \citep{34_attention-models-survey_2021}. 
In Figure \ref{fig:attention_architecture}(b), the attention function $g\left(\cdot\right)$ consists of two key components including an alignment function and a distribution function, while the $f\left(\cdot\right)$ is used as a weighted sum function to calculate the final attention value.

\begin{table}[!htb]
\caption{Summary of Common Alignment Functions.}
\renewcommand{\arraystretch}{1.5}
\centering
\begin{tabular*}{1 \linewidth}[c]{ccc}
\toprule
Function & Equation & Description \\
\midrule
Cos similarity & $Cos(K,Q)$ & $Cos\left(\cdot\right)$: Cosine similarity \\
Dot product & $Q^TK$ & $T$: Matrix transpose \\
Scaled dot product & $\frac{Q^TK}{\sqrt{{dim}_K}}$ & ${dim}_K$: Dimension of $K$ \\
General & $Q^T\Theta K$ & $\Theta$: trainable parameters \\
Biased general & $Q^T\Theta K+b$ & $b$: biase \\
Additive & $\omega^T\ Act(\Theta_1Q+\Theta_2K+b)$ & $\omega$: trainable parameters \\
Concat & $\omega^T\ Act(\Theta[Q:K]+b)$ & $Act$:activate function \\
\bottomrule
\label{tab:alignment_func}
\end{tabular*}
\end{table}

The alignment function is a process of calculating the attention alignment score, which is the core of the attention mechanism \citep{34_attention-models-survey_2021}. There are many classical alignment functions in Table \ref{tab:alignment_func}, defined as:

\begin{equation}
    scores=Sim\left(Q,K\right)
\label{eq:eq_alignment_score}
\end{equation}

The distribution function converts the attention score to the attention coefficients $\alpha$. We define the distribution function in a unified form as below:

\begin{equation}
    \alpha=Norm\left(scores\right)
\label{eq:eq_attention_coefficient_norm}
\end{equation}
where, $Norm(\cdot)$ refers to a distribution function. The softmax function $softmax(\cdot)$ is the most commonly used distribution function, defined as:

\begin{equation}
    softmax\left(x_i\right)=\frac{exp(x_i)}{\sum_{n=1}^{N}{exp(x_n)}}
\label{eq:eq_softmax}
\end{equation}
where, $x_i \in X$ represents the $i$-th element in $X$ and $N$ is the total number of elements in $X$. $exp(\cdot)$ is an exponential function. With a softmax function, the calculated attention score is normalized to the probability distribution in $\left[0, 1\right]$ and sum to $1$.

Finally, an aggregation process for the attended representations with a weighted sum function to get the attention value:

\begin{equation}
    Z=f\left(X\right)=\alpha\ V
\label{eq:eq_attention_mechanisms}
\end{equation}

Almost all existing attention mechanisms can be written into the above formulations under the generalized framework.

When an attention mechanism is used to compute a representation of a single sequence, it is commonly referred to as self-attention or intra-attention \citep{35_tran-survey_2020}. Generally, the complete self-attention layer (SAL) with scaled dot product can be expressed as:

\begin{equation}
    Z=SAL(Q,K,V)=softmax\left(\frac{QK^T}{\sqrt{{dim}_Q}}\right)V
\label{eq:eq_sal}
\end{equation}

The self-attention mechanism calculates the scaled dot product of the query with all the keys to find the similarity scores between them, and this product is normalized by the softmax function for obtaining the attention weight. The weighted sum function is used to get the attention value $Z$ as outputs. Finally, the whole process of self-attention is shown in Figure \ref{fig:attention_architecture}(b). Further, we rewrite the self-attention layer under the generalized framework:

\begin{equation}
    Q,K,V=Linear\left(X\right)
\label{eq:eq_sal_linear_transformation}
\end{equation}
\begin{equation}
    scores=\frac{QK^T}{\sqrt{{dim}_Q}}
\label{eq:eq_sal_score}
\end{equation}
\begin{equation}
    \alpha=softmax\left(scores\right)
\label{eq:eq_sal_a}
\end{equation}
\begin{equation}
    Z=\alpha\ V
\label{eq:eq_sal_z}
\end{equation}
where $Linear\left(\cdot\right)$ refers to a linear transformation, $Q^T$ is a matrix transpose operation. $Q^TK$ is the dot product operation, which is widely used in self-attention to calculate the attention score.

The attention mechanism in GNNs allows the neural networks to learn a dynamic and adaptive aggregation of the neighborhood as well as enables the model to avoid or ignore noisy parts of the graph \citep{36_AGNN_2018}. Based on attention functions, we define three types of attention mechanisms in GNNs including local attention, global attention, and feature fusion attention. 

\begin{figure}[!htbp]
    \centering
    \includegraphics[width=1\linewidth]{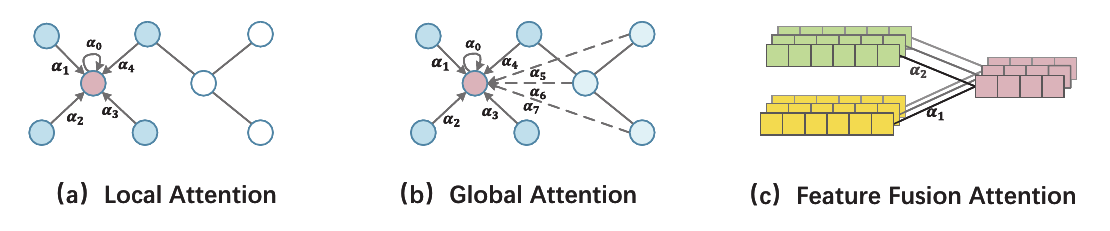}
    \caption{Different attention mechanisms in GNNs. Local attention focuses only on the direct neighbors of the central node, while global attention focuses on all nodes in the graph. Feature fusion attention is a weighted approach from the perspective of feature fusion using an attention mechanism.}
    \label{fig:gnn_attention_mechanisms}
\end{figure}

Local attention focuses on local neighbors, that is, directly or locally connected neighbors. GAT \citep{13_GAT_2018} learns different aggregating weights for each neighbor representation through local attention, as shown in Figure \ref{fig:gnn_attention_mechanisms}(a). Treating normalized attention coefficients as the relative weights between node pairs, attention-based GNNs aggregate and update the representations of nodes in the graph with a weighted sum function, then propagate them to higher layers \citep{12_diffusion_2019}. As shown in Figure \ref{fig:gnn_attention_mechanisms}(b), the self-attention mechanism in graph transformers can be viewed as passing messages among all nodes in the entire graph with global attention, regardless of the input graph connectivity \citep{37_PAGAT_2019}. Different from local attention and global attention, feature fusion attention directly adopts the traditional attention mechanisms, focusing on feature fusion beyond the node, as shown in Figure \ref{fig:gnn_attention_mechanisms}(c).

From the node view, let ${h}_i^l,{h}_j^l\in H^l$ refers to the representation of the node $v_i$ and $v_j$ in the $l^{th}\in L$ layer, and ${{h}_i^0={x}}_i^0$ as the input features of node $v_i$. 
As an initial step, a shared linear transformation with a weight matrix $\Theta^l$ is applied to every node in the graph. After the linear transformation, the alignment function obtains the attention alignment score, which indicates the importance of its local neighbors. 
In addition, the distribution function converts the attention score to the attention coefficient, which makes the attention coefficients comparable across different nodes. 
Finally, the weighted sum function is used to update and aggregate the representations of nodes in the graph. The above local attention layer can be defined as follows:

\begin{equation}
    {h}_i^l=\Theta^l\ {x}_i^l\ ,\ {h}_j^l=\Theta^l\ {x}_j^l 
\label{eq:eq_h_l}
\end{equation}
\begin{equation}
    {scores}_{ij}^l=Sim\left({h}_i^l,{h}_j^l\right)
\label{eq:eq_attention_score}
\end{equation}
\begin{equation}
    \alpha_{ij}^l=Norm\left({scores}^l\right)
\label{eq:eq_attention_coefficient}
\end{equation}
\begin{equation}
    {h}_i^{l+1}=\sigma\left\{\sum_{j\in\mathrm{\Gamma}_i}{\alpha_{ij}^l\ {h}_j^l}\right\}
\label{eq:eq_h+l+1}
\end{equation}
where $\sigma$ denotes the non-linear activation function. $\mathrm{\Gamma}_i$ refers to the local neighborhoods of the given node $v_i$, which is exactly the first-order node in GAT. $Sim(\cdot)$ represents the alignment functions (listed in Table \ref{tab:alignment_func}) and $Norm(\cdot)$ is the distribution functions, such as softmax, and sigmoid. The softmax function is defined in Equation \ref{eq:eq_softmax} and the sigmoid function is defined as:

\begin{equation}
    sigmoid\left(x\right)=\frac{1}{1+exp(-x)}
\label{eq:eq_sigmoid}
\end{equation}

The alignment function in GAT with a single-layer feed-forward neural network parameterized by $\omega^T\in\mathbb{R}^{2\ dim}$ denotes as:

\begin{equation}
    {scores}_{ij}^l=LeakyReLU\left(\omega^T\left[{h}_i^l\Vert{h}_j^l\right]\right)
\label{eq:eq_gat_alignment_func}
\end{equation}
where $\left[\cdot\Vert\cdot\right]$ refers to the operation of vector concatenation and $LeakyReLU(\cdot)$ represents an activation function. Compared with $ReLU(\cdot)$, $LeakyReLU(\cdot)$ introduces a constant parameter $\lambda\in\left(0,1\right)$, defined as:
\begin{equation}
    LeakyReLU\left(x\right)=\left\{
    \begin{aligned}
    x & , & if \  x > 0  \\
    \lambda x & , & if \  x \leq 0
    \end{aligned}
    \right.
\label{eq:eq_LeakyReLU}
\end{equation}

The other widely used alignment function is a dot-product operation in self-attention, especially in Graph Transformers. The self-attention mechanism linearly projects the center node feature ${h}_i^l$ to get the query vector and project the neighboring node features ${h}_j^l$ to get the key and value vectors. The dot-product operation from the node view can be expressed as:

\begin{equation}
    {scores}_{ij}^l={{h}_i^l}^T{h}_j^l\ 
\label{eq:eq_dot_product}
\end{equation}

To stabilize the learning process of self-attention, multi-head self-attention is also introduced into attention-based GNNs. While the multi-head attention aggregator can explore multiple representation subspaces between the center node and its neighborhoods \citep{28_GATv2_2021}. Every self-attention block contains its queries, keys, values, and learnable weight matrices, as shown in Figure \ref{fig:attention_architecture}(c). We summarize the two forms of multi-head self-attention blocks: concatenated multi-head and averaging multi-head. Multi-head self-attention (MHSA) with a concatenate operation can be formulated as:

\begin{equation}
    {h}_i^{l+1}=MHSA\left({head}_1,{head}_2,\cdots,{head}_K\right)={\Vert}_{k=1}^K\sigma\left(\sum_{j\in\mathrm{\Gamma}_i}{\alpha_{ij}^{lk}\ \Theta^{lk}{h}_j^{lk}}\right)
    \label{eq:eq_MHSA}
\end{equation}
where $\Vert$ represents concatenation, $\alpha_{ij}^{lk}$ refers to the normalized attention coefficients computed by the $k^{th}$ attention head in the $l^{th}$ layer. $\sigma$ is the non-linear activation function. And the averaging multi-head can be formulated as:

\begin{equation}
    {h}_i^{l+1}=MHSA\left({head}_1,{head}_2,\cdots,{head}_K\right)=\sigma\left(\frac{1}{K}\sum_{k=1}^{K}\sum_{j\in\mathrm{\Gamma}_i}{\alpha_{ij}^{lk}\ \Theta^{lk}\ {h}_j^{lk}}\right)
\label{eq:eq_MHSA_average}
\end{equation}

Transformers are deep feed-forward artificial neural networks with a self-attention mechanism \citep{2022_survey_Transformer}. Similar to the traditional Transformer in Figure \ref{fig:attention_architecture}(d), Graph Transformers take multi-head self-attention as the core component. The Graph Transformer is introduced as a novel encoder-decoder architecture built with multiple blocks of self-attention, without convolution or recurrent modules  \citep{38_GTN_2019}. Each Transformer layer has two parts: multi-head self-attention modules and a position-wise feed-forward network (FFN)  \citep{39_tran-review_2022}. The generalized graph transformer with multi-head self-attention (MHSA) and feed-forward block (FFN) can be defined as:

\begin{equation}
    {z}^{l+1}=LayerNorm\left(MHSA\left({h}^l\right)\right)+{h}^l
\label{eq:eq_transformer_z}
\end{equation}
\begin{equation}
    {h}^{l+1}=LayerNorm\left(FFN\left({z}^{l+1}\right)\right)+{z}^{l+1} 
\label{eq:eq_transformer_h}
\end{equation}
where ${z}^{l+1}$ is the output of the first stage in the graph transformer layer. $LayerNorm\left(\cdot\right)$ is the layer normalization layer. ${h}^{l+1}$ is the final output of this transformer block, the representations of the node in $\left(l+1\right)^{th}$ layer. The FFN layer with a non-linear activation function can be defined as:

\begin{equation}
    FFN\left({h}^l\right)=\sigma\left({h}^l\ \Theta_1+b_1\right)\Theta_2+b_2
\label{eq:eq_transformer_ffn}
\end{equation}

\section{Taxonomy of Attention-based GNNs}
\label{Taxonomy}

\begin{table}[!htb]
\caption{The characteristics of different developmental stages}
\centering
\begin{tabular}{ccl}
\hline
\multicolumn{2}{c}{\textbf{Stages}} & \multicolumn{1}{c}{\textbf{Characteristics}} \\ \hline
\multicolumn{2}{c}{\textbf{GRANs}} & \begin{tabular}[c]{@{}l@{}}Local attention.\\ Limited by the long-term dependence bottleneck and order \\ problem of RNN.\end{tabular} \\ \hline
\multirow{2}{*}{\textbf{GATs}} & \textbf{Intra-layer} & \begin{tabular}[c]{@{}l@{}}Local attention.\\ Stacking layers leads to high complexity and over-smoothing.\end{tabular} \\ \cline{2-3} 
 & \textbf{Inter-layer} & \begin{tabular}[c]{@{}l@{}}Feature fusion attention.\\ Feature selection operation from different feature spaces, \\ such as different levels, different channels, different views,\\ or different time slices.\end{tabular} \\ \hline
\multicolumn{2}{c}{\textbf{\begin{tabular}[c]{@{}c@{}}Graph\\    Transformers\end{tabular}}} & \begin{tabular}[c]{@{}l@{}}Global attention.\\ Directly obtain high-order neighborhood information.\\ Catastrophic Spatio-temporal complexity and limited on \\ large-scale graphs.\end{tabular} \\ \hline
\end{tabular}
\label{tab:stage_characteristics}
\end{table}

\begin{figure}[!htp]
    \centering
    \includegraphics[width=1\linewidth]{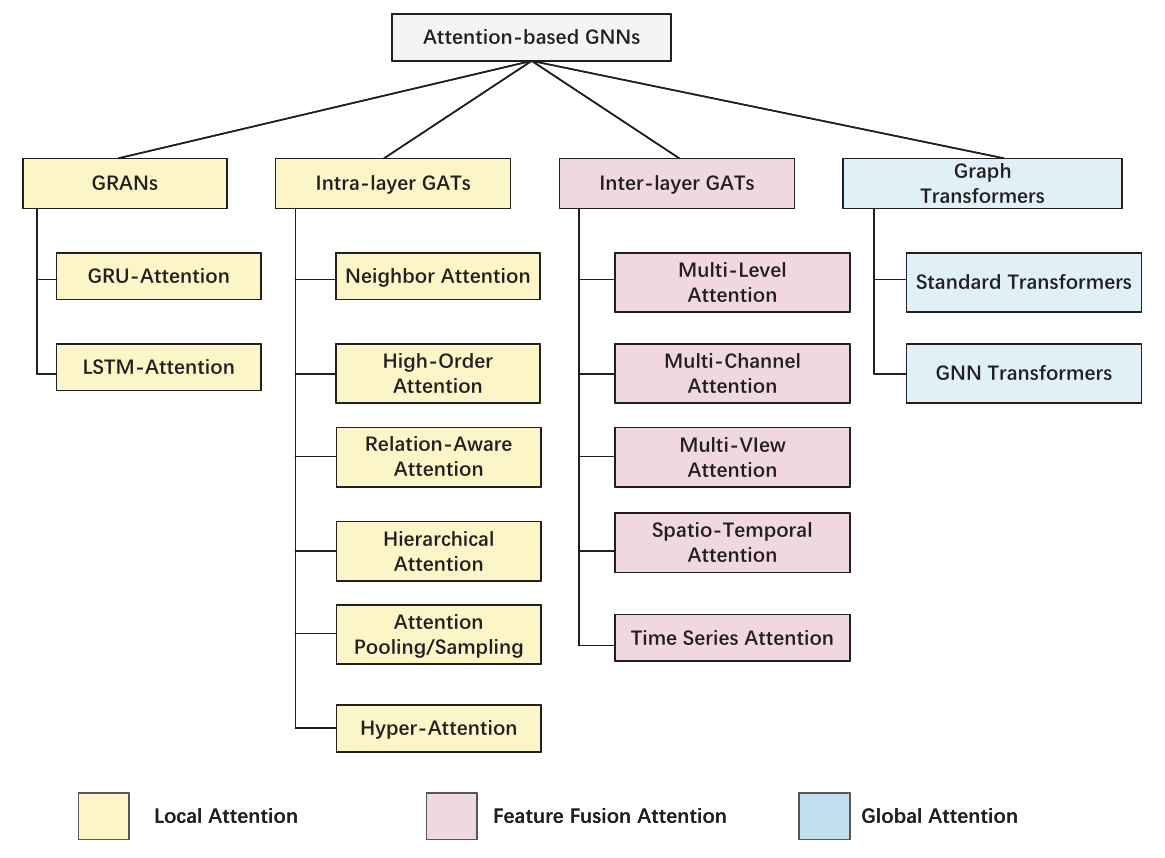}
    \caption{Classification breakdown of methods for attention-based GNNs}
    \label{fig:attention_GNNs_classification}
\end{figure}

In this section, we outline a novel two-level taxonomy for attention-based GNNs from the perspective of development history and architectural perspectives. The upper level reveals the three developmental stages of attention-based graph neural networks, including Graph Recurrent Attention Networks (GRANs), Graph Attention Networks (GATs: intra-layer and inter-layer), and Graph Transformers.
We summarize their characteristics in Table \ref{tab:stage_characteristics}. The lower level focuses on various typical architectures of each stage, as shown in Figure \ref{fig:attention_GNNs_classification}.

Figure \ref{fig:attention_GNNs_classification} shows the two-level taxonomy of attention-based GNNs. The  color in the figure represents the type of attention mechanism employed in a category.

GRANs: This kind of works focus on recurrent neural networks(RNNs) to learn representations of graph-structured data. 
GRANs are inspired by attention-based recurrent neural networks(RNNs) in deep learning  \citep{23_GRAN_2019}. 
Therefore, we naturally divide these methods into two subclasses: GRU-Attention \citep{22_GaAN_2018} and LSTM-Attention \citep{26_JK-Net_2018}.

\begin{sidewaystable}[!htbp]
\caption{Categories of each developmental stage and representative works.}
\centering
\scriptsize
\resizebox{\textwidth}{70mm}{
\begin{tabular}{ccc}
\hline
\multicolumn{2}{c}{\textbf{Categories}} & \textbf{Representative works} \\ \hline
\multirow{2}{*}{
\textbf{GRANs}} & 
\textbf{\begin{tabular}[c]{@{}c@{}}GRU-\\ Attention\end{tabular}} & \begin{tabular}[c]{@{}c@{}}GGNN \citep{21_GGNN_2016}, GRNN \citep{59_GRNN_2020}, GaAN \citep{22_GaAN_2018},\\ GRAN \citep{23_GRAN_2019} \end{tabular} \\ \cline{2-3} & 
\textbf{\begin{tabular}[c]{@{}c@{}}LSTM-\\ Attention\end{tabular}} & JK-Net \citep{26_JK-Net_2018}, GAM \citep{61_GAM_2018}, GeniePath \citep{62_GeniePath_2019} \\ \hline
\multirow{6}{*}{
\textbf{\begin{tabular}[c]{@{}c@{}}Intra-layer \\ GATs\end{tabular}}} & \textbf{\begin{tabular}[c]{@{}c@{}}Neighbor \\ Attention\end{tabular}} & \begin{tabular}[c]{@{}c@{}}GAT \citep{13_GAT_2018}, DMP \citep{64_MS-heterophily_2021}, SuperGAT \citep{19_SuperGAT_2021},\\ CPA \citep{2020_GATs_CPA}, GATv2 \citep{28_GATv2_2021}, PPRGAT \citep{65_pagerank_2021}, Simple-HGN \citep{3_Simple-HGN_2021}, \\ DPGAT \citep{28_GATv2_2021}, C-GAT \citep{27_C-GAT_2019}, GANet \citep{66_GANet_2019},\\ CAT \citep{67_CAT_2021}, MSNA \citep{68_MSNA_2020}, AGNN \citep{36_AGNN_2018},\\ HTNE \citep{69_HTNE_2018}, HAT \citep{70_HyperGAT_2021}, HGCN \citep{71_HGCN_2019},\\ Hype-HAN \citep{72_Hype-han_2021}, DKGAT \citep{73_DKGAT_2020}, DAGL \citep{74_DAGL_2021}, \\SCGA \citep{75_SCGA_2021}, Co-GAT \citep{76_Co-GAT_2021},  ED-GAT \citep{77_ED-GAT_2020},\\  TD-GAT \citep{78_TD-GAT_2019}, GATON \citep{79_GATON_2020}, GRAM \citep{80_GRAM_2017}\end{tabular} \\ \cline{2-3}  & 
\textbf{\begin{tabular}[c]{@{}c@{}}High-Order \\ Attention\end{tabular}} & \begin{tabular}[c]{@{}c@{}} SPAGAN \citep{40_SPAGAN_2019}, PaGNN \citep{81_PaGNN_2021}, CGAT \citep{82_CGAT_2020},\\ ADSF \citep{83_ADSF_2019}, MAGNA \citep{84_MAGNA_2021}, T-GAP \citep{85_T-GAP_2021} \end{tabular} \\ \cline{2-3} 
 & \textbf{\begin{tabular}[c]{@{}c@{}}Relation-Aware \\ Attention\end{tabular}} & \begin{tabular}[c]{@{}c@{}}SiGAT \citep{86_SiGAT_2019}, SNEA \citep{41_SNEA_2020}, RGAT \citep{42_RGAT_2019},\\ WRGNN \citep{87_WRGNN_2021}, HetSANN \citep{88_HetSANN_2020}, EAGCN \citep{89_EAGCN_2018},\\ TALP \citep{90_TALP_2020},  KGAT \citep{91_KGAT_2019}, GATNE \citep{92_GATNE_2019},\\ CGAT \citep{93_CGAT_2020}, RelGNN \citep{94_RelGNN_2021}, AFE \citep{95_AFE_2019},\\ DisenKGAT \citep{96_DisenKGAT_2021}, GTAN \citep{97_tri-attention_2021}, R-GAT \citep{98_R-GAT_2020},\\ ReGAT \citep{99_ReGAT_2019}, AD-GAT \citep{100_AD-GAT_2021} \end{tabular} \\ \cline{2-3} 
 & \textbf{\begin{tabular}[c]{@{}c@{}}Hierarchical \\ Attention\end{tabular}} & \begin{tabular}[c]{@{}c@{}}HAN \citep{43_HAN_2019}, PSHGAN \citep{101_PSHGAN_2022}, PRML \citep{102_PRML_2017},\\ GraphHAM \citep{44_GraphHAM_2022}, LAN \citep{104_LAN_2019}, \\ RGHAT \citep{105_RGHAT_2020}, DANSER \citep{106_DANSER_2019}, UVCAN \citep{107_UVCAN_2019},\\ GCATSL \citep{108_GCATSL_2021}, DAGC \citep{109_DAGC_2020}, AGCN \citep{110_AGCN_2020},\\ HGAT \citep{111_HGAT_2021} \end{tabular} \\ \cline{2-3} 
 & \textbf{\begin{tabular}[c]{@{}c@{}}Attention \\ Sampling/Pooling\end{tabular}} & \begin{tabular}[c]{@{}c@{}} GAW \citep{45_GAW_2018}, NLGCN \citep{112_NLGCN_2021}, SAGPool \citep{113_SAGPool_2019},\\ Attpool \citep{114_Attpool_2019}, ChebyGIN \citep{46_ChebyGIN_2019} \end{tabular}\\ \cline{2-3} 
 & \textbf{Hyper-Attention} & \begin{tabular}[c]{@{}c@{}} Hyper-SAGNN \citep{47_Hyper-SAGNN_2020}, HHGR \citep{115_HHGR_2021}, \\ HyperTeNet \citep{116_HyperTeNet_2021}, Hyper-GAT \citep{117_Hyper-GAT_2021} \end{tabular}\\ \hline
\multirow{5}{*}{\textbf{\begin{tabular}[c]{@{}c@{}}Inter-Layer \\ GATs\end{tabular}}} & \textbf{\begin{tabular}[c]{@{}c@{}}Multi-Level \\ Attention\end{tabular}} & DAGNN \citep{48_DAGNN_2020}, TDGNN \citep{129_TDGNN_2021}, GAMLP \citep{49_GAMLP_2022} \\ \cline{2-3} 
 & \textbf{\begin{tabular}[c]{@{}c@{}}Multi-Channel \\ Attention\end{tabular}} & FAGCN \citep{50_FAGCN_2021}, ACM \citep{51_ACM_2021} \\ \cline{2-3} 
 & \textbf{\begin{tabular}[c]{@{}c@{}}Multi-View \\ Attention\end{tabular}} & \begin{tabular}[c]{@{}c@{}}AM-GCN \citep{52_AM-GCN_2020}, MV-GNN \citep{53_MV-GNN_2021}, GENet \citep{130_GENet_2021},\\ UAG \citep{131_UAG_2021}, MVE \citep{132_MVE_2017} , MGAT \citep{133_MGAT_2020} \end{tabular} \\ \cline{2-3} 
 & \textbf{\begin{tabular}[c]{@{}c@{}}Spatio-Temporal\\ Attention\end{tabular}} & \begin{tabular}[c]{@{}c@{}}DySAT \citep{54_DySAT_2018}, TemporalGAT \citep{134_TemporalGAT_2020}, GAEN \citep{135_GAEN_2021},\\ MMDNE \citep{55_MMDNE_2019}, TGAT \citep{136_TGAT_2020},\\  T-GNN \citep{138_T-GNN_2022}, ST-GCN \citep{139_ST-GCN_2018}, GMAN \citep{140_GMAN_2020},\\ ASTGCN \citep{141_ASTGCN_2019}, ConSTGAT \citep{142_ConSTGAT_2020}\end{tabular} \\ \cline{2-3} 
 & \textbf{\begin{tabular}[c]{@{}c@{}}Time Series \\ Attention\end{tabular}} & RainDrop \citep{56_RainDrop_2021}, MTAD-GAT \citep{57_MTAD-GAT_2020}, GACNN \citep{143_GACNN_2020} \\ \hline
\multirow{2}{*}{\textbf{\begin{tabular}[c]{@{}c@{}}Graph \\ Transformers\end{tabular}}} & \textbf{\begin{tabular}[c]{@{}c@{}}Standard \\ Transformers\end{tabular}} & \begin{tabular}[c]{@{}c@{}}Graphormer \citep{5_Graphormer_2021}, HOT \citep{30_transformers-generalize_2021},\\ PAGAT \citep{37_PAGAT_2019}, GTA \citep{152_GTA_2021}, GT \citep{153_GT_2020},\\ SAN \citep{154_SAN_2021}, GraphBert \citep{24_Graph-bert_2020}, UniMP \citep{155_UniMP_2021}\end{tabular} \\ \cline{2-3} 
 & \textbf{\begin{tabular}[c]{@{}c@{}}GNN \\ Transformers\end{tabular}} & \begin{tabular}[c]{@{}c@{}}GROVER \citep{156_GROVER_2020}, UGformer \citep{25_UGformer_2019}, GMT \citep{157_GMT_2021},\\ GTN \citep{38_GTN_2019}, GraphFormers \citep{158_GraphFormers_2021}, HGT \citep{159_HGT_2020},\\  GTOS \citep{160_GTOS_2020}, GraphWriter \citep{161_GraphWriter_2019}, KHGT \citep{162_KHGT_2021},\\ GATE \citep{163_GATE_2021}, STAGIN \citep{164_STAGIN_2021}\end{tabular} \\ \hline
\end{tabular}
}
\label{tab:works}
\end{sidewaystable}

Intra-layer GATs: This kind of works introduce the attention mechanism into the local neighborhoods in the single-layer neural network with local attention. 
The intra-layer GATs \citep{13_GAT_2018} usually place the local attention on the local neighborhoods in the graph within a single-layer neural network. Intra-layer GATs can be further divided into six subclasses, namely neighbor attention \citep{13_GAT_2018}, high-order attention \citep{40_SPAGAN_2019}, relation-aware attention \citep{41_SNEA_2020,42_RGAT_2019}, hierarchical attention \citep{43_HAN_2019,44_GraphHAM_2022}, attention sampling/pooling \citep{45_GAW_2018,46_ChebyGIN_2019}, and hyper-attention \citep{47_Hyper-SAGNN_2020}. 
Specifically, neighbor attention only considers the direct neighborhoods, while high-order attention takes $k^{th}$-hop neighborhoods and subgraphs into consideration. 
Relation-aware attention takes into account different types of relations. 
In addition to focusing on node-level attention, hierarchical attention also considers higher-level attention, such as path, group, and relationship. 
Attention sampling/pooling refers to attention-based GNNs used for sampling or pooling. Hyper-attention represents special attention mechanisms for hypergraphs.

Inter-layer GATs: This kind of works usually select features beyond neural network layers with multiple feature spaces, not just local neighborhoods. 
Across the neural network layer, attention in inter-layer GATs can be regarded as an operation of cross-layer fusion of different feature spaces with feature fusion attention.
In this term, attention-based GNNs dynamically select features from different levels, different channels, different views, or different time slices. 
Therefore, we further divide these methods into five sub-categories (i.e., multi-level attention \citep{48_DAGNN_2020,49_GAMLP_2022}, multi-channel \citep{50_FAGCN_2021,51_ACM_2021}, multi-view \citep{52_AM-GCN_2020,53_MV-GNN_2021}, Spatio-temporal attention \citep{54_DySAT_2018,55_MMDNE_2019}, and time series attention \citep{56_RainDrop_2021,57_MTAD-GAT_2020}). 
By considering temporal attributes, Spatio-temporal attention usually uses time, spatial attention, or both in dynamic graphs, while time-series attention needs to construct dynamic graphs from time-series data first.

Graph Transformers: In the past two years, Transformers \citep{58_transformer_survey_2021} have achieved superior performance in many tasks of NLP, CV, and GRL.
Graph Transformers generalize the Transformer architecture to graph representation learning, capturing long-range dependency \citep{5_Graphormer_2021}.
Different from previous methods with local attention, Graph Transformers learn higher-order graph properties directly via global attention. 
Graph Transformers have developed rapidly in the field of graph deep learning, especially in the task of graph classification on small and medium-sized graphs. 
We further divide Graph Transformers into two sub-categories, namely standard Transformers \citep{5_Graphormer_2021} and GNN Transformers \citep{25_UGformer_2019}. Standard Transformers usually utilize the self-attention mechanism to all nodes of the input graph, ignoring adjacencies between nodes, while GNN Transformers use the GNN layer to obtain adjacency information. 

We tabularize representative works for each sub-category in the different stages as shown in Table \ref{tab:works}. In the following sections, we detail each representative work in terms of motivation, characteristics, or functions

\section{GRANs}
\label{GRANs}

GRANs are inspired by attention-based recurrent neural networks (RNNs). Table \ref{tab:works} shows the representative works of two kinds of GRANs, namely GRU-Attention and LSTM-Attention.

\subsection{GRU-Attention}
Gated Recurrent Units (GRU) can be regarded as operators acting on the current input and previous state to control how much of the input should be taken into account, and how much past information should be remembered (or forgotten) in the computation of the new state \citep{59_GRNN_2020}. 

Based on the previous work on GNNs \citep{63_GNN-model_2008}, GGNN introduces GRU with a soft attention mechanism. 
The graph representation in GGNN uses context to focus attention on which nodes are important to the current graph-level task \citep{21_GGNN_2016}. 
Further, GRNN proposes a general learning framework leveraging the notion of a recurrent hidden state together with graph signal processing (GSP) \citep{59_GRNN_2020}. 
Under this framework, GRNN generates meaningful representations of graph signals by incorporating the importance of a node’s features to its neighbors with three different gating mechanisms: time, node, and edge gates.

Unlike the traditional multi-head attention mechanism, which equally computes all attention heads, GaAN uses a convolutional sub-network to control each attention head’s importance \citep{22_GaAN_2018}. 
GaAN assigns different importance to each head through computing an additional soft gate mechanism between 0 and 1, 0 for low importance and 1 for high importance \citep{22_GaAN_2018}. 
Similar to the graph generation task in GaAN, GRAN captures the auto-regressive conditioning between the already-generated and to-be-generated parts of the graph using GNNs with attention, which not only reduces the dependency on node ordering but also bypasses the long-term bottleneck caused by the sequential nature of RNNs \citep{23_GRAN_2019}. 

\subsection{LSTM-Attention}
Another classic and simple RNN structure is Long Short-Term Memory (LSTM). 
GraphSAGE \citep{16_inductive_2017} first introduces the LSTM in GNNs as an aggregator, different from mean and pooling aggregators. 
Cooperating with sampling, GraphSAGE aggregates the messages from the local neighborhood of the node through the aggregator. 
In the same way, JK-Net \citep{26_JK-Net_2018} adapts LSTM-Attention as a layer-aggregation to aggregate the jumping representations from the previous layers. 
LSTM-Attention in JK-Net is node adaptive because the attention scores are different for each node \citep{26_JK-Net_2018}.

Due to the ability to capture long-range dependencies, GAM adopts an LSTM and proposes a solution for graph classification based on attention-guided walks \citep{61_GAM_2018}. 
The attention mechanism in GAM focuses on small but informative parts of the graph and avoids noise in the rest of the graph. 
GeniePath proposes an adaptive path layer with two complementary functions: breadth function and depth function \citep{62_GeniePath_2019}. 
The adaptive breadth function learns the importance of different sized neighborhoods and adaptively selects a set of important 1-hop neighbors with a parameterized generalized linear attention operator. 
While the adaptive depth function for depth exploration can extract useful signals and filter noisy signals up to long-distance neighbors with a gated unit. 
Even though LSTMs have a more sophisticated memory model when compared to simple RNNs, it has been shown that they still have trouble remembering information that was inputted too far in the past \citep{61_GAM_2018}.

\section{Intra-Layer GATs}
\label{Intra-Layer GATs}

Considering different local neighborhoods and different functions, intra-layer GATs can be further divided into six subclasses including neighbor attention, high-order attention, relation-aware attention, hierarchical attention, attention sampling/pooling, and hyper-attention. 
Table \ref{tab:works} shows the representative works of each subclass.

\subsection{Neighbor Attention}
As the most famous attention network, GATs are widely used in a variety of graph-structured data scenes. 
GATs with neighbor attention compute the hidden representations of each node in the graph by attending to its neighbors in the local neighborhoods. 
GAT takes the lead in introducing the attention mechanism into graph neural networks to aggregate representations of local neighbor nodes in the graph \citep{13_GAT_2018}.
Every node in GAT attends to the neighbors in its neighborhood and treats its representation as a query \citep{28_GATv2_2021}. 
GAT takes a single-layer feed-forward neural network (FFN) as an alignment function to calculate the attention scores:

\begin{equation}
{scores}_{ij}^l=LeakyReLU\left(\omega^T\left[{\Theta\ {h}}_i^l\Vert\Theta{h}_j^l\right]\right)
\label{eq:27}
\end{equation}
where ${scores}_{ij}^l$ indicates the importance of the node $v_j$ to node $v_i$ in the $l^{th}$ layer. 
As an initial step, the FFN adopts a shared linear transformation, parametrized by a weight matrix $\Theta$, to perform feature transformation for every node in the graph. 
With a learnable weight vector $\omega\in\mathbb{R}^{2\ dim}$, the FNN applies the LeakyReLU as a non-linear activation function. 
Ignoring unconnected node pairs, GAT only computes the masked attention for each node with its first-order neighbors via local attention, as shown in Figure \ref{fig:gnn_attention_mechanisms}(a). 
Subsequently, GAT normalizes the attention scores using the softmax function:

\begin{equation}
\alpha_{ij}={softmax}_j\left({scores}_{ij}\right)=\frac{exp\left({scores}_{ij}\right)}{\sum_{k\in\mathrm{\Gamma}_i} e x p\left({scores}_{ik}\right)}
\label{eq:28}
\end{equation}
where $v_j$ and $v_k$ are the neighbor nodes in the neighborhood $\mathrm{\Gamma}_i$ of node $v_i$. 
Different from the non-negative distribution function via a softmax non-linear function in GAT, DMP \citep{64_MS-heterophily_2021} takes tanh as a non-linear function, whose output is zero-centered and ranged in $\left(-1,\ 1\right)$, defined as:

\begin{equation}
\alpha_{ij}=tanh\left({scores}_{ij}\right)=\frac{exp\left({scores}_{ij}\right)-exp\left({-scores}_{ij}\right)}{exp\left({scores}_{ij}\right)+exp\left(-{scores}_{ij}\right)}
\label{eq:29}
\end{equation}

In DMP, the positive weights correspond to low-passing filtering capturing the similarity between nodes, while the negative weights facilitate the filtering of high-frequency to reduce the difference \citep{50_FAGCN_2021,64_MS-heterophily_2021}.

In addition, GAT adapts the multi-head attention to stabilize the learning process of self-attention. 
A concatenated multi-head concatenates the output representations of different heads. 
An averaging multi-head is employed to average the representations of different heads in the final layer as output for the prediction task.

The attention function in GAT always weighs one key at least as much as any other key, unconditioned on the query \citep{28_GATv2_2021}. 
To address this limitation, GATv2 \citep{28_GATv2_2021} makes a simple fix by modifying the order of operations in the alignment function, defined as:

\begin{equation}
{scores}_{ij}^l=\omega^T\ LeakyReLU\left(\Theta\left[{\ {h}}_i^l\Vert{h}_j^l\right]\right)
\label{eq:30}
\end{equation}

GATv2 performs an additional empirical comparison to DPGAT \citep{28_GATv2_2021}, applying the scaled dot-product attention of the Transformer \citep{118_transformer_2017}: 

\begin{equation}
{scores}_{ij}^l=\frac{{{({h}}_i^l}^T\Theta_Q){({{h}_j^l}^T\Theta_K)}^T}{\sqrt{dim}}
\label{eq:31}
\end{equation}

PPRGAT \citep{65_pagerank_2021} incorporates the Personalized PageRank (PPR \citep{119_GNN-PR_2018}) information into the GAT and GATv2, which utilizes the full potential of GATs. 
For this, the alignment functions in GAT and GATv2 are modified as:

\begin{equation}
{scores}_{ij}^l=LeakyReLU\left(\omega^T\left[{\Theta\ {h}}_i^l\Vert\Theta{h}_j^l\Vert\Pi_{ij}\right]\right)
\label{eq:32}
\end{equation}
\begin{equation}
{scores}_{ij}^l=\omega^T\ LeakyReLU\left(\Theta\left[{\ {h}}_i^l\Vert{h}_j^l\Vert\Pi_{ij}\right]\right)
\label{eq:33}
\end{equation}
where $\Pi_{ij}$ is the PPR matrix in the preprocess step before starting the training. 
Simple-HGN \citep{3_Simple-HGN_2021} also adopts the same method for feature concatenation with the edge relationship type between the node $v_i$ and $v_j$.

Instead of the soft graph attention operator (GAO) in GAT, GANet \citep{66_GANet_2019} introduces the hard attention operator (HGAO) and channel-wise graph attention operator (CGAO) with dot product attention. 
HGAO uses the hard attention mechanism by attending to only important nodes, while CGAO avoids the dependency on the adjacency matrix, leading to dramatic reductions in computational resource requirements \citep{66_GANet_2019}. 
Besides considering the layer-wise node features propagated within the GNN, CAT learns representations of nodes in the graph with conjoint attention, considering additional structural information, such as node cluster embedding and higher-order structural correlations when computing attention scores \citep{67_CAT_2021}. 
MSNA \citep{68_MSNA_2020} extracts comprehensive and expressive neighborhood features with two neighborhood attention including self neighborhood attention network (SNAN) and cross neighborhood attention network (CNAN). 
The SNAN predicts the link of two nodes by encoding and matching their respective neighborhood information, while the CNAN with a cross neighborhood attention directly captures structural interactions between two nodes \citep{68_MSNA_2020}.

To alleviate the over-fitting problem, C-GAT \citep{27_C-GAT_2019} proposes two additional margin-based constraints as loss functions on GAT. 
CPA preserves cardinality information in attention-based aggregation, which can be applied to any kind of attention mechanisms \citep{2020_GATs_CPA}.
To improve the graph attention model for noisy graphs, SuperGAT  \citep{19_SuperGAT_2021} exploits two commonly used attention mechanisms: a single-layer feed-forward neural network (FFN) in GAT and a dot product (DP), for a self-supervised task to predict edges \citep{19_SuperGAT_2021}.

Apart from FNN and dot product, there are a series of classical attention-based GNNs adopting other alignment functions to calculate attention scores, such as cosine similarity and Euclidean distance. 
AGNN \citep{36_AGNN_2018} removes all the intermediate fully-connected layers and replaces the propagation layers with an attention mechanism that computes attention with cosine similarity. 
HTNE chooses a negative Euclidean distance function to score the affinity between the source and history node to better determine the influence of historical neighbors on the current neighbors of a node \citep{69_HTNE_2018}.

Beyond Euclidean space, HAT \citep{70_HyperGAT_2021} attempts the GAT with an attention mechanism in hyperbolic space \citep{120_HyperbolicGNN-review_2022}. 
To calculate the attention, HAT transforms the features in a graph into gyrovector space and then proposes the attention-based hyperbolic proximity to aggregate the features \citep{70_HyperGAT_2021}. 
HGCN also introduces a hyperbolic attention-based aggregation scheme based on the Riemannian manifold, to learn inductive node representations for hierarchical and scale-free graphs \citep{71_HGCN_2019}. 
Hyperbolic attention operation makes use of hyperbolic geometry in both the computation of the attention weights and the aggregation operation  \citep{121_Hyper-Att_2018}. 
Further, Hype-HAN is a hierarchical embedding method based on three types of hyperbolic manifolds on Riemannian geometries \citep{120_HyperbolicGNN-review_2022}, including the Lorentz model, Klein model, and Poincaré model, for text classification tasks  \citep{72_Hype-han_2021}.

Up to the present, a great variety of graph attention networks and their variants have been proposed in many artificial intelligence fields, such as natural language processing \citep{76_Co-GAT_2021,77_ED-GAT_2020,78_TD-GAT_2019,79_GATON_2020}, computer vision \citep{73_DKGAT_2020,74_DAGL_2021,75_SCGA_2021}, and medical health \citep{80_GRAM_2017}. In the field of natural language processing, we can think of words as nodes and then construct graph models. GNNs with neighbor attention utilize the dependency relationship among words for sentiment analysis \citep{76_Co-GAT_2021,77_ED-GAT_2020,78_TD-GAT_2019}, and topic modeling \citep{79_GATON_2020}. GRAM proposes a graph-based attention model for healthcare representation learning \citep{80_GRAM_2017}. For computer vision, many works have extended neighbor attention to extracting local features in images or videos, e.g., 3D object recognition \citep{73_DKGAT_2020}, image restoration \citep{74_DAGL_2021}, and video-grounded dialogue \citep{75_SCGA_2021}. 

\subsection{High-Order Attention}
\begin{figure}[!htbp]
    \centering
    \includegraphics[width=0.9\linewidth]{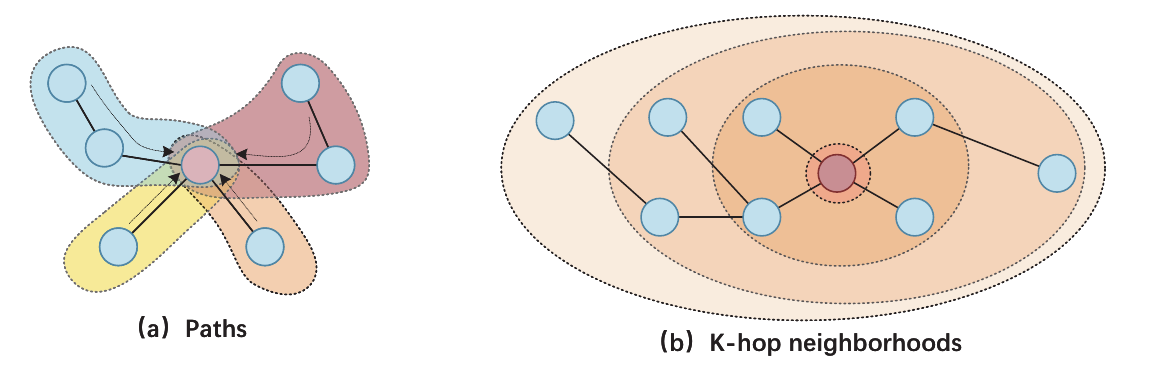}
    \caption{Higher-order message-passing along paths and $k$-hop neighborhoods. Path-based message-passing provides a mechanism for aggregating messages along meta-paths, while $k$-hop neighborhoods aggregate messages from different hop neighborhoods.}
    \label{fig:7}
\end{figure}

Neighbor attention often only focuses on aggregating information from first-order neighbors within each layer. To obtain the representations of higher-order neighbors, the conventional approach is stacking multiple layers, which also brings high complexity and over-smoothing \citep{40_SPAGAN_2019}. Such an attention mechanism is limited because it does not consider nodes that are not connected by an edge, which could provide important contextual information.
In Figure \ref{fig:7}, high-order attention (path-based and $k$-hop neighbors) aggregates information from distant neighbors and explores the graph topology, all in a single neural network layer\citep{40_SPAGAN_2019}.

To account for higher-order neighbors, one can stack multiple layers to enlarge the size of the receptive field, horribly bringing high complexity. 
The other is high-order neighbors in a single neural network layer, achieved by shortest paths, to capture the more global graph topology. 
Unlike conventional neighbor attention that carries out node-based attention within each layer, SPAGAN proposes path-based attention accounting for the influence of a sequence of nodes, or shortest path, between the current node and its higher-order neighbors \citep{40_SPAGAN_2019}. 
PaGNN \citep{81_PaGNN_2021} develops a novel path-aware GNN for link prediction, integrating inter-action and neighborhood information via broadcasting and aggregating operations. 
To effectively preserve the structural topology and semantic properties in heterogeneous information networks, CGAT \citep{82_CGAT_2020} adopt multiple meta-paths-based sampling and pre-training process with pair-wise attention.

To fully exploit rich, high-order structural details in GATs, ADSF \citep{83_ADSF_2019} proposes an adaptive structural fingerprints model. 
The key idea of ADSP is to update the representation of each node within a local receptive field consisting of its high-order neighbors. 
Modeling attention flow on graphs is another way to obtain higher-order information, which effectively contributes to the information flow implemented through message passing \citep{122_MAF_2018}. 
To eliminate noisy high-frequency information from the graph, MAGNA \citep{84_MAGNA_2021} captures large-scale structural information in each layer with a low-pass filter. 
MAGNA propagates the attention scores across the graph with Personalized PageRank, increasing the receptive field for each layer of the GNN \citep{84_MAGNA_2021}. 
Similarly, T-GAP models a path traversal with the soft approximation of attention flow, iteratively propagating the attention value of each node to its outgoing neighbor nodes \citep{85_T-GAP_2021}.

\subsection{Relation-Aware Attention}

\begin{figure}[!htbp]
    \centering
    \includegraphics[width=0.8\linewidth]{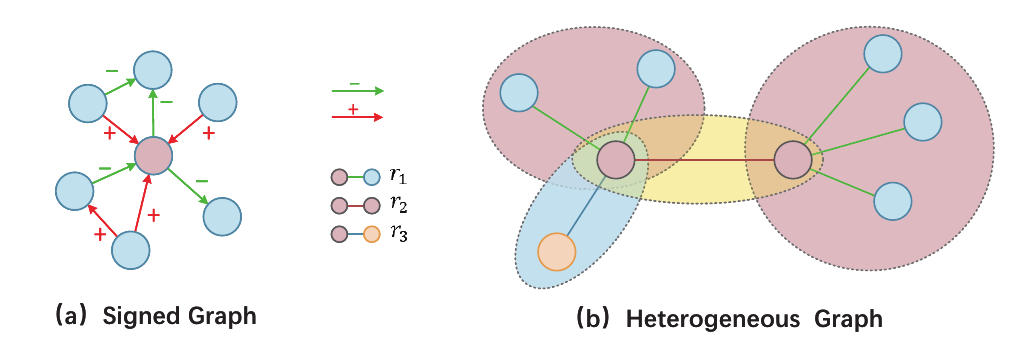}
    \caption{Relation-aware neighborhoods in signed graph and heterogeneous graph. (a) There are two types of edges in the signed graph: positive and negative. (b) Different color blocks represent different types of relationships.}
    \label{fig:8}
\end{figure}

Most of the interactions in social networks are positive relationships, such as friendship, following, and support. 
Meanwhile, some negative links exist in the real world indicating disapproval, disagreement, or distrust  \citep{86_SiGAT_2019}. 
However, some of the graphs are signed graphs with both positive and negative links in Figure \ref{fig:8}(a). 
GAT is designed to graph only considering positive links and ignoring the negative ones in the signed graph. 
Signed Graph Attention Networks (SiGAT) generalizes GAT to signed graphs  \citep{86_SiGAT_2019}, which incorporates graph motifs into GAT to capture two famous theories in the signed graph, i.e., balance and status theory.
In SiGAT, different motifs represent different relation-aware influences, when aggregating and propagating messages on the signed graph to generate node embeddings via an attention mechanism \citep{86_SiGAT_2019}. 
Also in the signed graph, SNEA computes the importance coefficient for pair nodes connected by different types of links when aggregating the embedding with a self-attention mechanism \citep{41_SNEA_2020}.

More complex than signed networks, heterogeneous information networks (HINs) are also ubiquitous in our daily life. 
As shown in Figure \ref{fig:8}(b), HINs usually consist of various types of vertices connected by various types of relations \citep{123_Heterophily-survey_2022,124_heter-survey_2022}. 
Taking into consideration edge type $r\in R$, the attention function on HINs to calculate the attention scores can be defined as:

\begin{equation}
{scores}_{ij}^r=LeakyReLU\left(\omega^T\left[\Theta_{r}{\ {h}}_i^r\ \Vert\Theta_{r}{\ {h}}_j^r\right]\right)
\label{eq:34}
\end{equation}
\begin{equation}
\alpha_{ij}^r=\frac{exp\left({scores}_{ij}^r\right)}{\sum_{k\in\mathrm{\Gamma}_i^r} e x p\left({scores}_{ik}^r\right)}
\label{eq:35}
\end{equation}
\begin{equation}
{h}_i^{l+1}=\sigma\left\{\sum_{r\in R}\sum_{j\in\mathrm{\Gamma}_i^r}{\alpha_{ij}^r\ {\Theta_{r}\ {h}}_j^l}\right\}
\label{eq:36}
\end{equation}
where $r\in R$ is the type of edge, $\mathrm{\Gamma}_i^r$ refers to the neighborhood of the node $v_i$ with the edge type $r$.

Different relations convey distinct pieces of information. Relation-aware graph attention networks aggregate information with a masked self-attention that takes account of local relational structure as well as node features.
Considering the relation between nodes in HINs, RGAT proposes two variants,  Within-Relation Graph Attention (WIRGAT) and Across-Relation Graph Attention (ARGAT), under an additive and multiplicative logit construction \citep{42_RGAT_2019}. 
HetSANN  \citep{88_HetSANN_2020} aggregates multi-relational information in neighborhoods with two kinds of attention scoring functions. 
In HetSANN, the type-aware attention layer adopts the voices-sharing product. 
WRGAT \citep{87_WRGNN_2021} transforms the input graph into a computation graph containing both proximity and structural information with the different types of edges. 
The generated multi-relational graph has a high-level assortativity and preserves rich structural information from the original graph \citep{87_WRGNN_2021}. 
EAGCN proposes an edge attention-based multi-relational GCN on chemical graphs with multiple relationships \citep{89_EAGCN_2018}. 
To avoid the fusion vector ignoring the effects of the local type information, TALP models the effect of type information and fusion information from local and global perspectives simultaneously \citep{90_TALP_2020}, based on a two-layer graph attention architecture. 
KGAT adaptively aggregates the embeddings from neighbors of the current node and updates the node’s representation with an attention mechanism to distinguish the importance of the neighbors \citep{91_KGAT_2019}. 
To capture the influential information in different edge types, GATE formalizes the attributed multiplex heterogeneous network embedding problem as well as uses the attention mechanism \citep{92_GATNE_2019}. 

In particular, RelGNN \citep{94_RelGNN_2021} generates the states of different relations and leverages them along with the node states to weigh the messages. 
Further, RelGNN balances the importance of attribute features and topological features via a self-attention mechanism and then generates the final representations of nodes \citep{94_RelGNN_2021}. 
To learn both node and topic embeddings while preserving the graph structural information, CGAT \citep{93_CGAT_2020} guides information aggregation via a channel-aware attention mechanism focusing on the edges. 
Relational graph attention networks play a significant role in some specific scenarios with rich relational structure information, such as knowledge graph \citep{95_AFE_2019,96_DisenKGAT_2021,97_tri-attention_2021}, NLP \citep{98_R-GAT_2020}, CV \citep{99_ReGAT_2019}, and stock prediction \citep{100_AD-GAT_2021}. Since knowledge graphs are inherently graph-structured data with multiple relationship types, relation-aware attention is well-suited for tasks in knowledge graphs, such as relation prediction \citep{95_AFE_2019}, knowledge graph completion \citep{96_DisenKGAT_2021}, and question answering \citep{97_tri-attention_2021}.

\subsection{Hierarchical Attention}

\begin{figure}[!htbp]
    \centering
    \includegraphics[width=0.8\linewidth]{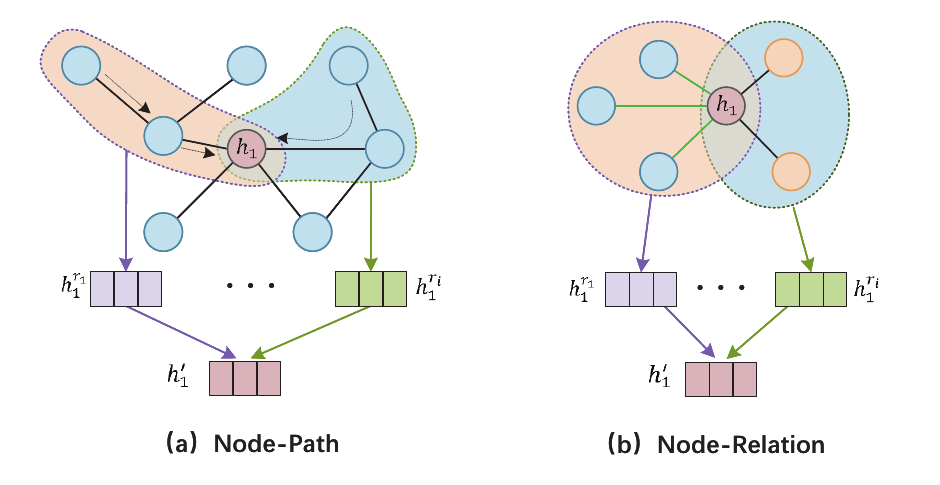}
    \caption{Hierarchical attention with node-path and node-relation. In addition to node-level attention, hierarchical attention often has higher-level attention, such as path, and relation.}
    \label{fig:9}
\end{figure}

Heterogeneous information networks (HINs) contain rich semantic information and hierarchical information, such as different types of nodes and links \citep{124_heter-survey_2022}. 
In Figure \ref{fig:9}(a), different meta-paths in HINs may extract diverse semantic information, while the relation-based hierarchical attention is illustrated in Figure \ref{fig:9}(b). 
Based on meta-paths, HAN \citep{43_HAN_2019} designs a heterogeneous graph neural network with hierarchical attention including node-level and semantic-level attention. 
The node-level attention learns the importance between a node and its meta-path-based neighbors, while the semantic-level attention learns the importance of different meta-paths \citep{43_HAN_2019}. 
Then HAN generates node representations by aggregating the received information from meta-path-based neighbors in a hierarchical manner. 
Similarly, PSHGAN \citep{101_PSHGAN_2022} first learns the weights of two nodes in the meta-path or meta-structure via a local attention mechanism. 
Then, PSHGAN learns a global attention weight based on meta-paths and meta-structures. 
In the end, PSHGAN uses computed dual-level attention to aggregate and update the representations of nodes. 
PRML \citep{102_PRML_2017} focuses on both node-level and path-level attention proximity of the endpoints based on their betweenness paths to discriminate the representations. 

RGHAT, equipped with a hierarchical attention mechanism including entity- and relation-level attention, can effectively aggregate the local neighborhood information of each entity \citep{105_RGHAT_2020}. 
The entity-level attention highlights the importance of different neighboring entities under the same relation, while the relation-level attention is inspired by the intuition that different relations have different weights for indicating an entity \citep{105_RGHAT_2020}. 
LAN aggregates neighbor information with both rules- and network-based attention weights \citep{104_LAN_2019} and focuses on both neighborhood features and query relations. 
GraphHAM \citep{44_GraphHAM_2022} aggregates neighboring states to generate node embeddings with group-level and individual-level attention for graph embedding. 
GraphHAM captures the information from neighborhoods of different scopes by stacking multiple layers, where the node states output by a lower layer are used as input to the layer above it \citep{44_GraphHAM_2022}.

Hierarchical attention-based GNNs are widely used in recommender systems  \citep{106_DANSER_2019,107_UVCAN_2019}, synthetic lethality \citep{108_GCATSL_2021}, point cloud \citep{109_DAGC_2020}, action detection \citep{110_AGCN_2020}, and short text classification \citep{111_HGAT_2021}.

\subsection{Attention Sampling/Pooling}

\begin{figure}[!htbp]
    \centering
    \includegraphics[width=0.8\linewidth]{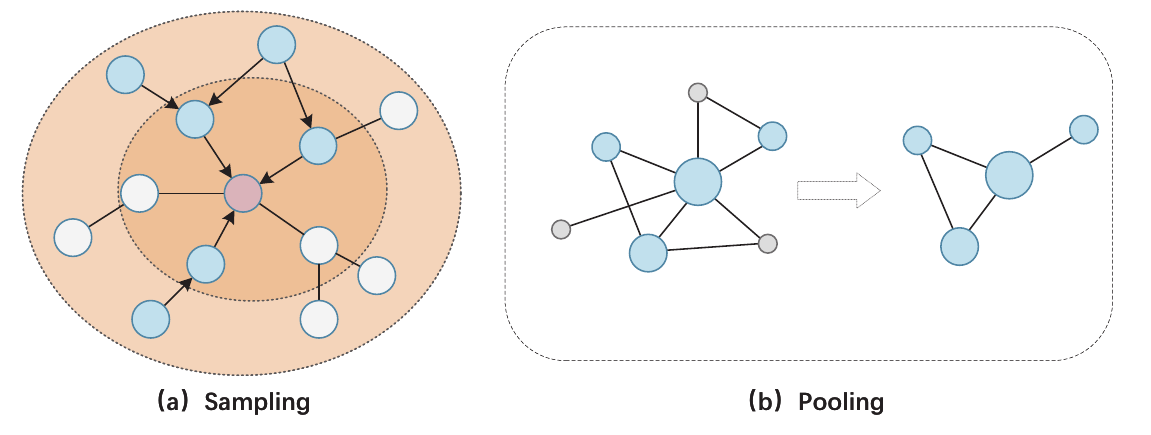}
    \caption{Sampling and Pooling in GNNs. Sampling is often used when messages are aggregated, while pooling is often used for readout operations for graph classification tasks.}
    \label{fig:10}
\end{figure}

Grap sampling is an effective method to select representative nodes from a large number of nodes \citep{16_inductive_2017}, as shown in Figure \ref{fig:10}(a). 
That is often adopted for large-scale graphs to improve efficiency \citep{125_GraphSAINT_2019}. 
GraphSAGE \citep{16_inductive_2017} first learns an aggregator function that generates embeddings by sampling and aggregating features from a node’s local neighborhood. 
Taking account of the attention mechanism, GAW \citep{45_GAW_2018} guides the random walk to optimize an upstream objective via the proposed attention model on the power series of the transition matrix. 
Furthermore, the attention mechanism in GAW only guides the random walk with a softmax function under the learning procedure \citep{45_GAW_2018}. 
To distinguish the importance of nodes, NLGAT \citep{112_NLGCN_2021} proposes a simple yet effective non-local aggregation framework with efficient attention-guided sorting for GNNs.

To generalize GNNs to larger, more complex, or noisy graphs, graph pooling aims to remove any number of nodes, so that the receives smaller graph turns into an input graph in the following layer, such as DiffPool \citep{126_hierarchical_2018}. 
For attention-based graph pooling in Figure \ref{fig:10}(b), the attention mechanism in the local neighborhood could distinguish the nodes that should be dropped and the nodes that should be retained. 
SAGPool \citep{113_SAGPool_2019} is a graph pooling method based on self-attention. 
Self-attention in SAGPool allows the proposed pooling method to consider both node features and graph topology \citep{113_SAGPool_2019}. 
To learn hierarchical representation for graph embedding, AttPool \citep{114_Attpool_2019} selects nodes that are significant for graph-level representation adaptively and then generates hierarchical features via the attention-based aggregator. 
ChebyGIN designs simple graph reasoning tasks to study the attention in GNNs under a controlled environment, and results show that attention can make GNNs more robust to larger and noisy graphs \citep{46_ChebyGIN_2019}.

\subsection{Hyper-Attention}

\begin{figure}[!htbp]
    \centering
    \includegraphics[width=0.6\linewidth]{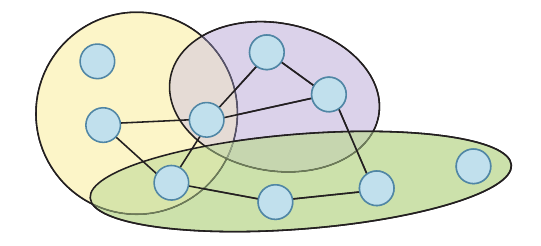}
    \caption{HyperGraph. Different color blocks represent different Hyper-edges. Hyper-edges are the interactions that occur between multiple nodes.}
    \label{fig:11}
\end{figure}

Modeling complex relationships between objects, hypergraphs exploit the high-order relationship and local clustering structure by hyperedges beyond a pairwise formulation \citep{127_hypergraph_2022}, as shown in Figure \ref{fig:11}. 
Recently, Hypergraphs with high-order relationships and variable hyperedges, attract much attention from researchers focusing on graph representation learning \citep{128_learnable-hypergraph_2022}.

Considering the complex relationships in hypergraphs, Hyper-SAGNN \citep{47_Hyper-SAGNN_2020} develops a new GNN model for the graph representation learning of general hypergraphs with various hyperedges. 
Since group members have different importance, HHGR  \citep{115_HHGR_2021} adopts a weighted sum function to generate the attentive group representation under a double-scale self-supervised setting. 
The hierarchical hypergraph convolutional network in HHGR consists of an attention-based group aggregator, which captures the user interactions within and beyond groups by propagating information from the user level to the group level \citep{115_HHGR_2021}. 
To learn the multi-hop relationship among the nodes in hypergraphs, HyperTeNet \citep{116_HyperTeNet_2021} designs a self-attention-based hypergraph neural network to learn the ternary relationships among the interacting nodes in a 3-uniform hypergraph. 
Hyper-GAT enhances the capacity of representation learning in high-order encoded by hyperedges with an attention module \citep{117_Hyper-GAT_2021}.

\section{Inter-Layer GATs}
\label{Inter-Layer GATs}

Across the neural network layer, inter-layer GATs combine representations from different feature spaces via feature fusion methods. According to different fusion methods, we divide these attention-based GNNs into five sub-categories, including multi-level attention, multi-channel attention, multi-view attention, Spatio-temporal attention, and time series attention, as listed in Table \ref{tab:works}.

\subsection{Multi-Level Attention}

\begin{figure}[!htbp]
\centering
\includegraphics[width=1.0\linewidth]{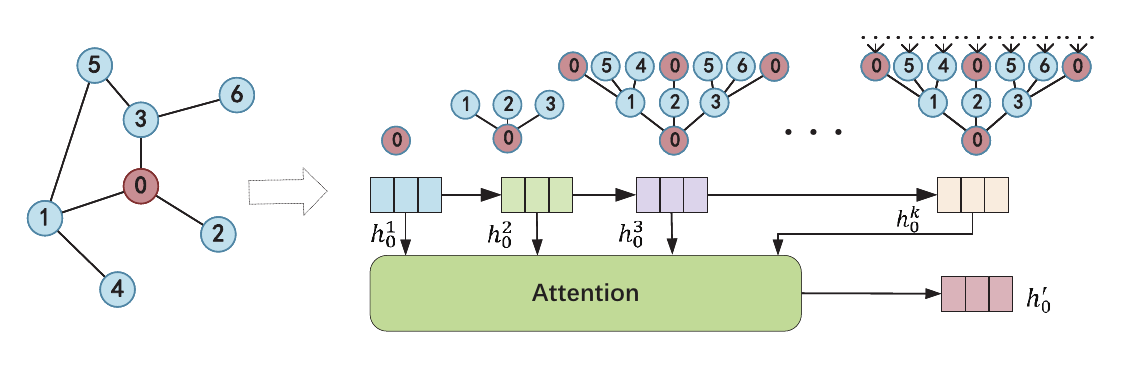}
\caption{Multi-level attention with different receptive fields. For central node $v_0$, its final representation is a combination of representations from different levels of neighborhoods through an attention mechanism.}
\label{Fig.12}
\end{figure}

Through stacking neural network layers, GNNs could learn node representations by aggregating the information from the multi-hop neighborhoods \citep{49_GAMLP_2022}. One layer of neighborhood aggregation in GNNs only considers immediate neighbors, but the performance decreases when going deeper to enable larger receptive fields due to over-smoothing \citep{48_DAGNN_2020}. Repeated propagation and aggregation make node representations indistinguishable from different classes \citep{144_Over-smoothing_2018}. To overcome the limitation of over-smoothing, some work has attempted to improve GNNs \citep{26_JK-Net_2018,112_NLGCN_2021,144_Over-smoothing_2018}. Among them, some works adaptively select multi-level features cross-layer with an attention mechanism, as shown in Figure \ref{Fig.12}.

DAGNN \citep{48_DAGNN_2020} transforms and propagates the representations of nodes with the ability to capture information from large and adaptive receptive fields. Specifically, DAGNN decouples transformation and propagation to leverage large receptive fields with multi-hop neighborhoods. For each node, DAGNN balances the information from local and global neighborhoods with an attention mechanism, thus leading to more discriminative node representations \citep{48_DAGNN_2020}. To alleviate feature smoothing among different layers, TDGNN \citep{129_TDGNN_2021} further disentangles neighborhoods in different layers with a tree decomposition process. TDGNN flexibly aggregates information from large receptive fields utilizing a deeper multi-hop dependency with graph diffusion \citep{129_TDGNN_2021}. To better combine the propagated features, GAMLP \citep{49_GAMLP_2022} adopts three receptive field attention mechanisms including smoothing attention, recursive attention, and JK attention.

\subsection{Multi-Channel Attention}

\begin{figure}[!htbp]
\centering
\includegraphics[width=0.9\linewidth]{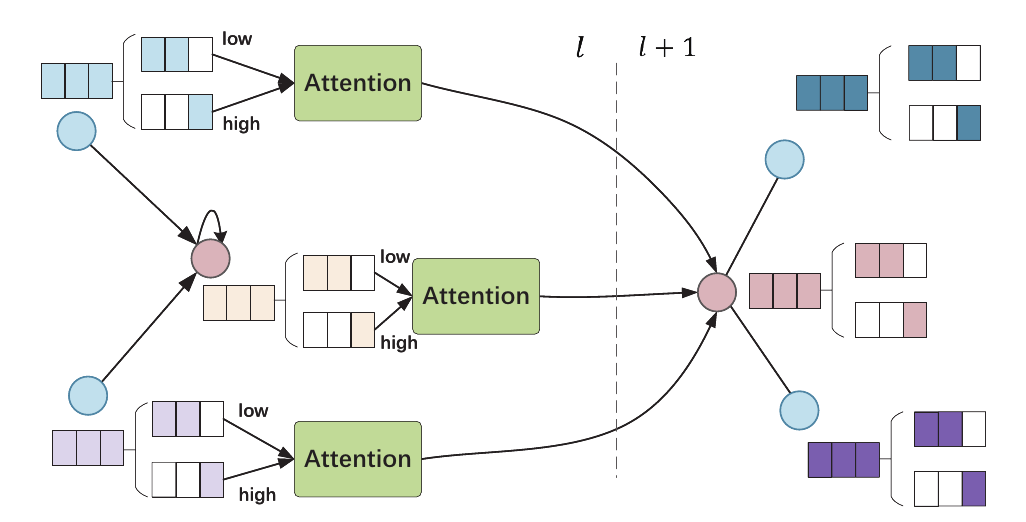}
\caption{Multi-channel attention with high- and low-frequency signals. Left: Low- and high-frequency information is aggregated by attention-weighted. Right: Update the representation of the central node based on messages from neighborhoods and itself in the upper layer.}
\label{Fig.13}
\end{figure}

Generally, most existing GNNs can be seen as low-pass filter, that updates node representations by aggregating information from neighbors in neighborhoods  \citep{145_SGC_2019}. In the low-pass filter, GNNs make use of node features as low-frequency signals, which are based on the assortative assumption that the node tends to connect with similar nodes. Sometimes, the useful high-frequency signals that capture the difference between nodes are mixed or ignored \citep{50_FAGCN_2021}. Different nodes in the graph may have different needs for the information in the different channels (frequency) \citep{51_ACM_2021}, as shown in Figure \ref{Fig.13}.

FAGCN \citep{50_FAGCN_2021} adaptively aggregates signals with different frequencies during message passing. FAGCN first designs an experimental study about low-frequency and high-frequency signals, where the results show that exploring low-frequency signals only is distant from learning an effective node representation in different scenarios \citep{50_FAGCN_2021}. Based on this observation, FAGCN presents a novel frequency adaptation graph convolutional network via an attention mechanism to adaptively combine the low-frequency and high-frequency signals \citep{50_FAGCN_2021}. Also considering different signals, ACM \citep{51_ACM_2021} proposes a framework to adaptively exploit aggregation, diversification, and identity channels to address harmful heterophily. To adaptively aggregate information from different channels, ACM learns node-wise attention to combine the different signals in three channels \citep{51_ACM_2021}.

\subsection{Multi-View Attention}

\begin{figure}[!htbp]
\centering
\includegraphics[width=0.9\linewidth]{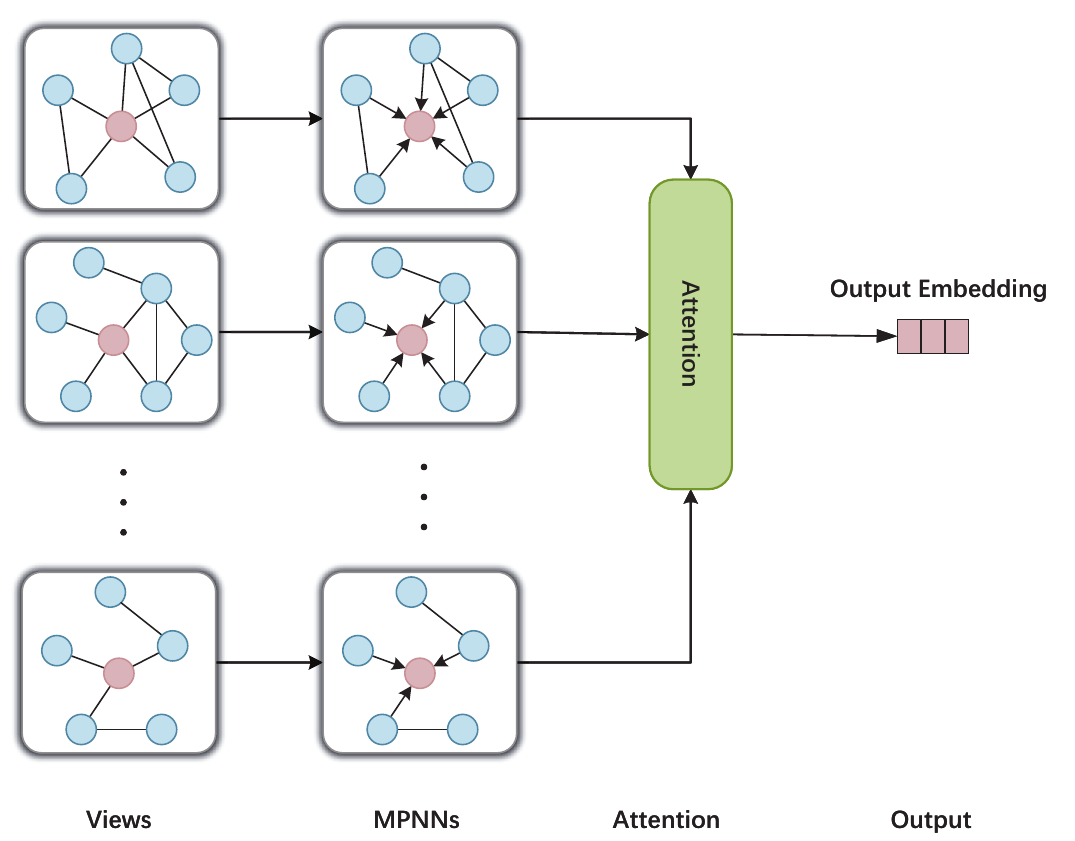}
\caption{Multi-View attention. Construct different graph structures for message passing and then use attention for message fusion.}
\label{Fig.14}
\end{figure}

Multi-view graph representation learning learns representations of nodes in the graphs with multiple views, such as different topologies, which aims to generate more robust node representations from different views \citep{53_MV-GNN_2021}. The multi-view GNNs usually integrate the embeddings from multiple feature spaces of different views to update the final node representations, and not all views aggregator subspaces are equally important. As shown in Figure \ref{Fig.14}, the attention mechanism in multi-view GNNs learns the adaptive importance weights of the embeddings from different views, to aggregate the view-specific node representations on each view \citep{53_MV-GNN_2021}.
 
To combine topological structures and node features substantially, AM-GCN \citep{52_AM-GCN_2020} selects the specific and common representations from different views including topological structures, node features, and their combinations simultaneously. AM-GCN first constructs a new graph topology based on features, named feature graph. With the feature graph and the topology graph, AM-GCN adaptively learns the deep correlation information in the feature space, topology space, and both of them, via an attention mechanism \citep{52_AM-GCN_2020}. UAG develops a Bayesian Uncertainty Technique (BUT) to explicitly capture uncertainties in GNNs and further employs an Uncertainty-aware Attention Technique (UAT) \citep{131_UAG_2021}. The UAT in UAG defends the adversarial attack on GNNs by assigning less impact on nodes with high uncertainty, thus, mitigating their impact on the final prediction \citep{131_UAG_2021}.

Based on the three views reflecting local, global structure, and feature similarity of nodes, MV-GCN \citep{53_MV-GNN_2021} designed an attention-based strategy to fuse the node representations from different views. GENet \citep{130_GENet_2021} introduces ensemble learning into GNNs based on different views with a drop-edge mechanism to construct subgraphs of different topological spaces \citep{146_DropEdge_2019}. Each member in the ensemble model generates individual embedding from different topology spaces and has a powerful capacity to resist noise perturbations in graph data \citep{130_GENet_2021}. MVE \citep{132_MVE_2017} promotes the collaboration of different views and lets them vote for robust representations with an attention mechanism. During the voting process in MVE, the attention mechanism enables each node to focus on the most important views. MGAT \citep{133_MGAT_2020} utilizes an attention-based architecture to learn node representations from different multi-view. To collaboratively integrate multiple types of relationships in different views, MGAT aggregates the view-wise node representations via view-focused attention \citep{147_MGAT_2020}.

\subsection{Spatio-Temporal Attention}

\begin{figure}[!htbp]
\centering
\includegraphics[width=0.9\linewidth]{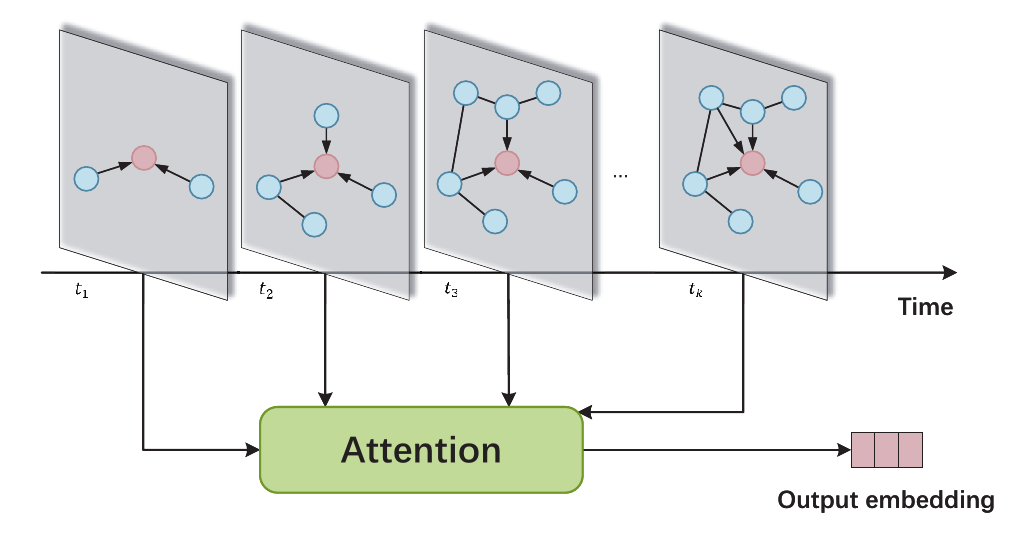}
\caption{Spatio-Temporal Attention in Dynamic Graph. Fusing messages from different time slices with attention mechanisms.}
\label{Fig.15}
\end{figure}

In the real world, some graph-structured data often show dynamic properties with continuously evolving network nodes and topology over time, named dynamic graphs \citep{148_Dynamic-survey_2020}. Compared to static graph learning, learning representations on dynamic graphs is challenging due to the temporal dependencies over time. For dynamic graph representation learning, we usually divide the dynamic graph into different snapshots, according to different time windows, as shown in Figure \ref{Fig.15}. Then, the dynamic graph representation learning learns low-dimensional node representations among a series of graph snapshots over the time step.

DySAT \citep{54_DySAT_2018} learns node representations that capture both structural properties and temporal evolutionary patterns. In detail, DySAT generates node representations by jointly applying self-attention layers along two dimensions including structural neighborhood and temporal dynamics \citep{54_DySAT_2018}. TemporalGAT \citep{134_TemporalGAT_2020} employs the self-attention mechanism and structural neighborhoods over temporal dynamics with a temporal convolutional network. Considering the neighborhoods in different snapshots, TemporalGAT learns dynamic node representation via self-attention strategy without violating the ordering of the graph snapshots \citep{134_TemporalGAT_2020}.

For dynamic graph representation learning, GAEN \citep{135_GAEN_2021} provides an evolving graph attention network across different time points. In addition, GAEN allows attention weights to share and evolve across all temporal networks based on their respective topology discrepancies \citep{135_GAEN_2021}. From the perspective of micro-and macro-dynamics, MMDNE \citep{55_MMDNE_2019} designs a temporal attention point process to capture structural and temporal properties at a fine-grained level. The micro-dynamics introduce the formation process of network structures in detail, while the macro-dynamics indicate the evolution pattern of the network scale \citep{55_MMDNE_2019}. Considering continuous time in dynamic graphs, TGAT \citep{136_TGAT_2020} presents a time-aware graph attention network to aggregate temporal-topological neighborhood features for inductive representation learning on temporal graphs. To handle continuous time, TGAT proposes a theoretically-grounded functional time encoding with a self-attention mechanism \citep{136_TGAT_2020}. 
Spatio-temporal attention has been successful in many application scenarios that can be modeled as dynamic graphs \citep{138_T-GNN_2022, 139_ST-GCN_2018, 140_GMAN_2020, 141_ASTGCN_2019, 142_ConSTGAT_2020}. In CV, spatio-temporal attention is applied to modeling a dynamic graph from a series of images at different times, such as trajectory prediction \citep{138_T-GNN_2022}, and action recognition \citep{139_ST-GCN_2018}. Traffic prediction, including traffic flow forecasting \citep{140_GMAN_2020, 141_ASTGCN_2019} and travel time estimation \citep{142_ConSTGAT_2020}, is the most common application scenario of spatio-temporal attention because of the large amount of spatio-temporal traffic network data in our daily life.

\subsection{Time Series Attention}

\begin{figure}[!htbp]
\centering
\includegraphics[width=0.9\linewidth]{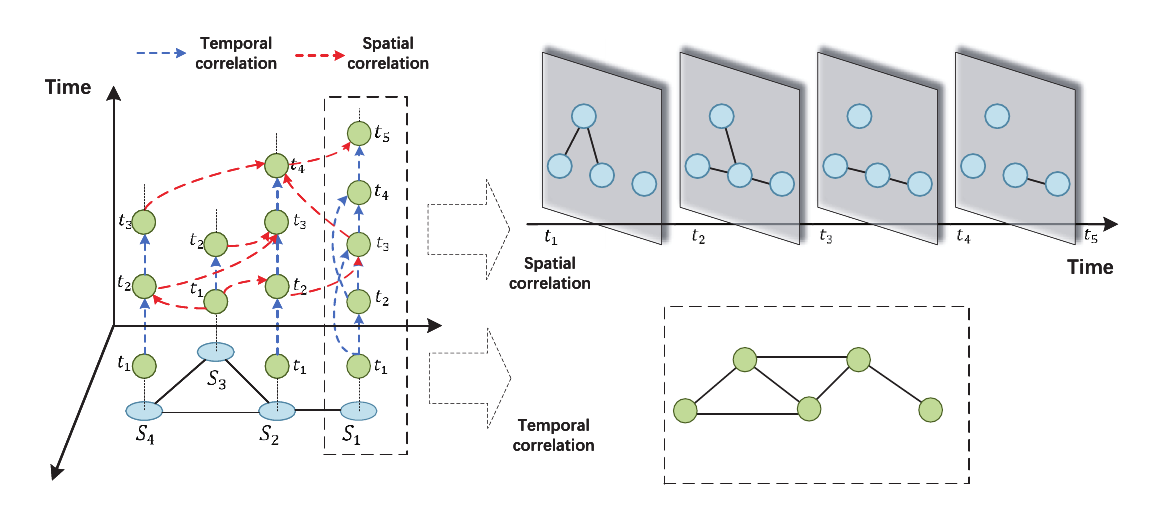}
\caption{Graph Attention in Time Series. Constructing graphs from a temporal and spatial perspective, modeling time series data with graphs.}
\label{Fig.16}
\end{figure}

A huge volume of data is generated overtime associated with various real-world systems. These data sets are often indexed by time, space, or both requiring appropriate approaches to analyze the data \citep{149_time-n_2021}. Time series analysis usually works on common tasks, including forecasting, anomaly detection, and classification \citep{150_tran-time-survey_2022}. If there are some association relationships between the time-series data that can be mined, graph neural networks can also play a role in time-series analysis. Numerous sensors have been deployed in different geospatial locations to continuously and cooperatively monitor the surrounding environment, such as climate science \citep{56_RainDrop_2021}. As shown in Figure \ref{Fig.16}, these sensors generate multiple geo-sensory time series, with spatial correlations between their readings \citep{151_GeoMAN_2018}.

To estimate the latent sensor graph structure and leverage the structure together with nearby observations, RainDrop \citep{56_RainDrop_2021} embeds irregularly sampled and multivariate time series while also learning the dynamics of sensors purely from observational data to predict misaligned readouts. For capturing time-varying dependencies among sensors, RainDrop can be interpreted as a graph neural network that sends messages over graphs with a self-attention mechanism \citep{56_RainDrop_2021}. To capture the relationships between different time series explicitly, MTAD-GAT \citep{57_MTAD-GAT_2020} considers each univariate time series as an individual feature and includes two graph attention layers in parallel to learn the complex dependencies of multivariate time series in both temporal and feature dimensions \citep{57_MTAD-GAT_2020}. Two parallel graph attention layers in MTAD-GAT learn the relationships between different time series and timestamps dynamically \citep{57_MTAD-GAT_2020}. GBikes \citep{143_GACNN_2020} proposes a data-driven Spatio-temporal Graph attention convolutional neural network (GACNN) for Bike station-level flow prediction. Based on the data-driven designs, GACNN predicts the fine-grained bike flows to/from each station, with attention mechanisms capturing and differentiating station-to-station correlations \citep{143_GACNN_2020}.

\section{Graph Transformers}
\label{Graph Transformers}

Taking into account whether the model uses the GNN layer to obtain adjacency information, we divide graph transformers into two sub-categories, i.e., standard Transformers and GNN Transformers. We enumerate the representative methods of standard Transformers and GNN Transformers in Table \ref{tab:works}.

\subsection{Standard Transformers}

Graph Transformers propagate node representations among all nodes in the graph regardless of whether two nodes are directly connected, while GAT focuses only on connected adjacent nodes. PAGAT \citep{37_PAGAT_2019} builds on longer-range dependencies in graph structure data and uses path features in molecular graphs to create a global attention layer.
 
\begin{figure}[!htb]
\centering
\includegraphics[width=0.9\linewidth]{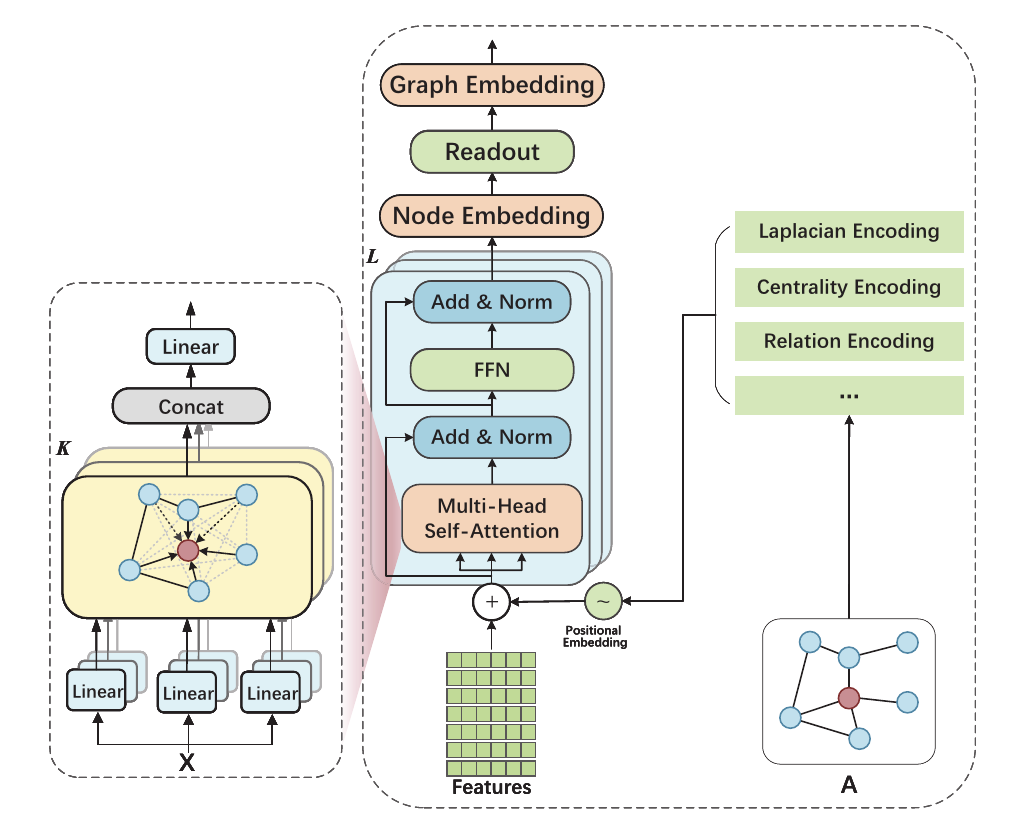}
\caption{The architecture of Graph Transformers. The attention mechanism in Graph Transformers can be seen as a fully connected GAT.}
\label{Fig.17}
\end{figure}

As shown in Figure \ref{Fig.17}, to utilize traditional Transformer architecture in the graph, it is necessary to define an effective position encoding with structural information of a graph, such as Laplacian encoding \citep{153_GT_2020}, relation encoding \citep{160_GTOS_2020}, centrality encoding, and edge encoding \citep{5_Graphormer_2021}. GT \citep{153_GT_2020} proposes a generalization of transformer networks to homogeneous graphs of the arbitrary structure via Laplacian eigenvectors as positional features. By leveraging the full spectrum of the Laplacian, SAN \citep{154_SAN_2021} can better detect similar sub-structures from their resonance with a fully-connected Transformer. Based on a batch of linkless subgraphs sampled from the original graph data, Graph-Bert \citep{24_Graph-bert_2020} learns the representations of the target node with the extended graph-transformer layers effectively. Graphormer \citep{5_Graphormer_2021} is also built upon the standard Transformer architecture for graph representation learning tasks. For the graph-level tasks, the positional encodings in Graphormer are three novel graph structural encodings including centrality encoding, spatial encoding, and edge encoding \citep{5_Graphormer_2021}.

Some research studies have explored graph representation learning using special mask mechanisms or higher-order Transformers. GTA \citep{152_GTA_2021} affords the advantages of both graph and sequence representations by encouraging the graph neural network characteristics of the transformer architecture. In addition, GTA takes the distance matrix of a molecular graph and atom-mapping matrix as mask self-attention and cross-attention. UniMP \citep{155_UniMP_2021} incorporates feature and label propagation at both training and inference time, taking feature embedding and masked label embedding as input information for propagation. Adopting the kernel attention approach to compute the pairwise weights, HOT \citep{30_transformers-generalize_2021} generalizes Transformers to any-order permutation invariant data involving sets, graphs, and hypergraphs.

\subsection{GNN Transformers}

\begin{figure}[!htbp]
\centering
\includegraphics[width=0.8\linewidth]{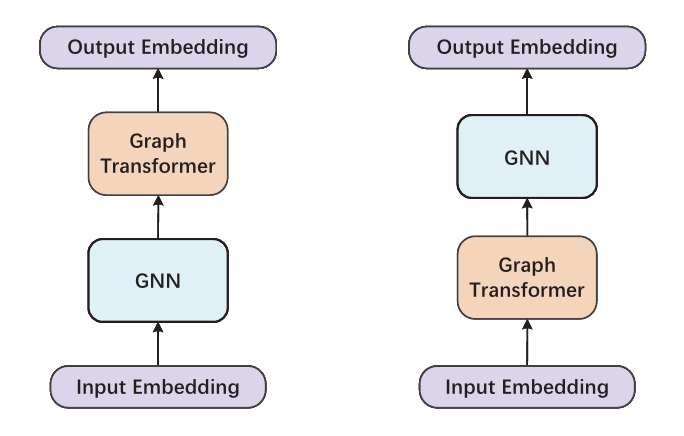}
\caption{The architecture of GNN Transformers. They can be seen as a combination of Graph Transformer and GNN.}
\label{Fig.18}
\end{figure}

Only utilizing the self-attention mechanism to all nodes of the input graph, Standard Transformers usually ignore the information about the neighborhood structure of the nodes \citep{25_UGformer_2019}. To overcome this limitation, GNN Transformers consist of Transformer layers with GNN layers. The GNN layer in GNN Transformers is usually serial (before or after) with the Transformer layer, as shown in Figure \ref{Fig.18}.
 
GraphFormers is a novel GNN Transformers architecture, where layerwise GNN components are nested alongside the Transformer blocks  \citep{158_GraphFormers_2021}. 
UGformer \citep{25_UGformer_2019} presents two graph transformer variants leveraging the transformer on a set of sampled neighbors or all input nodes. 
HGT \citep{159_HGT_2020} designs the heterogeneous mini-batch graph sampling algorithm for Web-scale heterogeneous graphs. 
To model heterogeneity, HGT designs node- and edge-type dependent parameters to characterize the heterogeneous attention over each edge, and maintain dedicated representations for different types of nodes and edges. To deliver a class of more expressive encoders of molecules, GROVER \citep{156_GROVER_2020} introduces the message-passing mechanism into the Transformer architecture.

In an end-to-end fashion, GTN \citep{38_GTN_2019} learns to transform a heterogeneous input graph into useful meta-path graphs for each task and learns node representation on the graphs. The Transformer layer in GTN learns a soft selection of edge types and composite relations for generating useful multi-hop connections so-called meta-paths \citep{38_GTN_2019}. To tackle the limitation of existing graph pooling methods, GMT \citep{157_GMT_2021} formulates the graph pooling problem as a multiset encoding problem with auxiliary information about the graph structure.

The Transformer model on graphs is more suitable for molecular graph tasks, molecular graph classification \citep{25_UGformer_2019}, molecular property prediction \citep{37_PAGAT_2019}, and retrosynthesis tasks \citep{152_GTA_2021}. Recently, graph transformers also stand out in other application scenarios, such as NLP \citep{160_GTOS_2020,161_GraphWriter_2019}, recommender systems \citep{162_KHGT_2021}, knowledge graphs \citep{163_GATE_2021}, and brain connections \citep{164_STAGIN_2021}.

\section{Comparison and Discussion}
\label{Comparison}

In this section, a model characteristics table is provided for a more comprehensive comparison. Each kind of model is characterized from nine perspectives, including local attention, high-order, global attention, multi-space, path-aware, edge-types, topology, times, and prior knowledge. In Table \ref{tab:comparison}, $\checkmark$ indicates that most models in the subclass have the corresponding characteristic. "local attention", "high-order" and "global attention" correspond to the information from neighborhoods that different models can obtain. "Multi-space" refers to the methods that adaptively select features from different feature spaces, resulting in a more robust latent representation. By considering the topological information of the graph, such as path, edge types, and topology, the model can learn a better representation that is more helpful for downstream tasks. These models are characterized by "path-aware", "edge-types" and "topology"; Time is also an important property when learning the representation of nodes, especially in dynamic graphs and time series. This kind of model is characterized by "time"; "prior knowledge" refers to the models that add prior knowledge in the process of graph representation learning.

\begin{sidewaystable}[!htbp]
\scriptsize
\caption{Characteristics of different models in subclasses from nine perspectives.}
\centering
\begin{tabular}{ccccccccccc}
\hline
\textbf{Stages} & \textbf{Subclasses} & \textbf{Local} & \textbf{\begin{tabular}[c]{@{}c@{}}High\\ Order\end{tabular}} & \textbf{Global} & \textbf{\begin{tabular}[c]{@{}c@{}}Multi\\ Space\end{tabular}} & \textbf{Path} & \textbf{\begin{tabular}[c]{@{}c@{}}Edge\\ Types\end{tabular}} & \textbf{Topology} & \textbf{Time} & \textbf{Prior} \\ \hline
\multirow{2}{*}{\textbf{GRANs}} & \textbf{\begin{tabular}[c]{@{}c@{}}GRU-\\ Attention\end{tabular}} & \checkmark &  &  &  &  & \checkmark &  &  &  \\
 & \textbf{\begin{tabular}[c]{@{}c@{}}LSTM-\\ Attention\end{tabular}} & \checkmark &  &  &  & \checkmark &  &  &  &  \\ \hline
\multirow{6}{*}{\textbf{\begin{tabular}[c]{@{}c@{}}Intra-Layer\\  GATs\end{tabular}}} & \textbf{\begin{tabular}[c]{@{}c@{}}Neighbor \\ Attention\end{tabular}} & \checkmark &  &  &  &  &  &  &  &  \\
 & \textbf{\begin{tabular}[c]{@{}c@{}}High-Order \\ Attention\end{tabular}} & \checkmark & \checkmark &  &  & \checkmark &  &  &  & \checkmark \\
 & \textbf{\begin{tabular}[c]{@{}c@{}}Relation-\\ Aware \\ Attention\end{tabular}} & \checkmark &  &  &  &  & \checkmark &  &  & \checkmark \\
 & \textbf{\begin{tabular}[c]{@{}c@{}}Hierarchical \\ Attention\end{tabular}} & \checkmark & \checkmark &  & \checkmark & \checkmark & \checkmark &  &  & \checkmark \\
 & \textbf{\begin{tabular}[c]{@{}c@{}}Attention \\ Sampling/Pooling\end{tabular}} & \checkmark &  &  &  &  &  & \checkmark &  &  \\
 & \textbf{\begin{tabular}[c]{@{}c@{}}Hyper-\\ Attention\end{tabular}} & \checkmark & \checkmark &  &  &  &  \checkmark &  &  & \checkmark \\ \hline
\multirow{5}{*}{\textbf{\begin{tabular}[c]{@{}c@{}}Inter-Layer\\  GATs\end{tabular}}} & \textbf{\begin{tabular}[c]{@{}c@{}}Multi-Level \\ Attention\end{tabular}} & \checkmark & \checkmark &  & \checkmark &  &  &  &  & \checkmark \\
 & \textbf{\begin{tabular}[c]{@{}c@{}}Multi-Channel \\ Attention\end{tabular}} &  &  &  & \checkmark &  &  &  &  & \checkmark \\
 & \textbf{\begin{tabular}[c]{@{}c@{}}Multi-View \\ Attention\end{tabular}} &  &  &  & \checkmark &  &  & \checkmark &  &  \\
 & \textbf{\begin{tabular}[c]{@{}c@{}}Spatio-Temporal \\ Attention\end{tabular}} &  &  &  & \checkmark &  &  &  & \checkmark & \checkmark \\
 & \textbf{\begin{tabular}[c]{@{}c@{}}Time Series\\ Attention\end{tabular}} &  &  &  & \checkmark &  &  & \checkmark & \checkmark &  \\ \hline
\multirow{2}{*}{\textbf{\begin{tabular}[c]{@{}c@{}}Graph \\ Transformers\end{tabular}}} & \textbf{\begin{tabular}[c]{@{}c@{}}Standard \\ Transformers\end{tabular}} &  & \checkmark & \checkmark &  & \checkmark & \checkmark & \checkmark &  & \checkmark \\
 & \textbf{\begin{tabular}[c]{@{}c@{}}GNN \\ Transformers\end{tabular}} & \checkmark & \checkmark & \checkmark &  & \checkmark & \checkmark & \checkmark &  & \checkmark \\ \hline
\end{tabular}
\label{tab:comparison}
\end{sidewaystable}

GRANs introduce RNN into graph representation learning, but at the same time, it is also limited by the inherent constraints of RNN \citep{23_GRAN_2019}. The original RNN architecture can only capture long-term dependencies on the ordered sequence, i.e. the output from one step will be used as input to the next step \citep{21_GGNN_2016}. However, the neighbors of the central node are disordered in graphs \citep{16_inductive_2017}. Although some methods deal with this problem by defining the order of nodes in advance \citep{16_inductive_2017}, different arrangement orders may result in different performances. On the other hand, GRU-Attention and LSTM-Attention are both based on local attention, which focuses on immediate neighbors. It is difficult to obtain information from distant neighbors. To learn long-range patterns in graphs,  GeniePath \citep{62_GeniePath_2019} introduces skip connections. Some models in GRU-Attention take into account the edge type of node pair \citep{23_GRAN_2019,59_GRNN_2020}, while LSTM-Attention models prefer path-based random walks \citep{61_GAM_2018,62_GeniePath_2019}.

Compared with GRANs, GATs can be calculated in parallel with the attention mechanism, because there is no need for sequential calculation. The attention mechanisms of GATs allow for dealing with variable-sized inputs, focusing on the most relevant parts of the input \citep{13_GAT_2018}.
Almost all models in intra-layer GATs aggregate and update the representation of nodes in defined local neighborhoods with local attention. 
GATs with local attention \citep{13_GAT_2018, 27_C-GAT_2019, 2020_GATs_CPA, 19_SuperGAT_2021, 28_GATv2_2021} enable specifying different weights to different nodes in the local neighborhood. However, GAT \citep{13_GAT_2018} also suffers from over-fitting and over-smoothing \citep{27_C-GAT_2019}. To address the above weaknesses, C-GAT improves GAT via margin-based constraints on attention during training \citep{27_C-GAT_2019}. To make a clear understanding of the discriminative capacities of GAT, CPA presents a theoretical analysis of the representational properties of the attention-based GNNs \citep{2020_GATs_CPA}. 
Considering the attribute homophily rate, DMP specifies every attribute propagation weight on each edge in graphs with heterophily \citep{64_MS-heterophily_2021}. To enable embeddings with much smaller distortion beyond Euclidean space, hyperbolic GATs \citep{71_HGCN_2019, 72_Hype-han_2021, 70_HyperGAT_2021} learn embeddings that preserve hierarchical structure in hyperbolic space. Unfortunately, It is difficult for these methods to directly obtain information from distant neighbors. To alleviate this limitation, GATs with high-order attention attempt to obtain information from distant neighbors through path-aware strategies \citep{40_SPAGAN_2019, 82_CGAT_2020, 81_PaGNN_2021}. 
To preserve the hierarchy in the graph topology, hierarchical attention defines two or more levels of attention mechanisms. Low-level attention is based on the node, while high-level attention can be based on paths \citep{111_HGAT_2021}, relations \citep{105_RGHAT_2020}, or groups \citep{44_GraphHAM_2022}. 
In such a case, hierarchical attention can learn specific hierarchical features of the graph. For graph classification, attention-based pooling \citep{114_Attpool_2019,113_SAGPool_2019} can learn hierarchical representations based on the attention mechanism. To apply GATs to complex graphs with multiple edge types, models with relation-aware attention usually define different relationships based on edge types \citep{86_SiGAT_2019, 88_HetSANN_2020, 94_RelGNN_2021}. While many complex graphs consist of relationships beyond pair-wise interactions, i.e., hyperedge that connects more than two nodes \citep{47_Hyper-SAGNN_2020}. Models with hyper-attention are designed for hypergraphs to extract patterns among higher-order interactions \citep{115_HHGR_2021}. 

Different from intra-layer GATs, inter-layer GATs extract useful hidden representations from different feature spaces via feature fusion attention, not just local neighborhoods. However, the performance of attention-based GNNs decreases when going deeper. Several recent studies attribute this performance deterioration to the entanglement of representation transformation and propagation in GNNs \citep{48_DAGNN_2020}. After decoupling these two operations, deeper GNNs can be used to learn node representations from larger receptive fields \citep{48_DAGNN_2020, 49_GAMLP_2022}. 
Then, multi-level attention adaptively fuses the representation of different-order neighbors to obtain high-order information. Similarly, multi-channel attention can distinguish signals with different frequencies during message passing \citep{50_FAGCN_2021}. To obtain the representations from different feature spaces, models with multi-view attention first construct several graphs with different topological structures based on the original graph and then fuse the representations of different views to obtain the final representation \citep{111_HGAT_2021, 133_MGAT_2020}. 
However, the above GATs ignore the graph with the time attribute. Spatio-Temporal attention and time-series attention can handle data with time attributes, such as dynamic graphs \citep{54_DySAT_2018, 134_TemporalGAT_2020, 136_TGAT_2020} and time series \citep{ 2021_GATs_mTAND, 56_RainDrop_2021}. Especially, time-series attention should first construct the graph structure from the time-series data. 

With the advent of Graph Transformers, GNNs do not need to obtain distant messages by stacking network layers. 
Graph Transformers can directly propagate the information among the nodes in the whole graph with global attention. Standard Transformers in graphs generate the position encoding with path-aware \citep{5_Graphormer_2021}, edge-types \citep{153_GT_2020}, or additional attributes \citep{154_SAN_2021, 24_Graph-bert_2020}. This kind of models are outstanding on the small molecular graph. However, Standard Transformers do not consider the original graph structure during training. To make up for this shortcoming, GNN Transformers jointly train the GNN layers and Transformers layers \citep{24_Graph-bert_2020}. This combination of models can effectively compensate for the shortcomings of the above single model. Further, we can capture both local and global information through a combination of local and global attention, which is difficult to achieve with a single model.

In addition, it is also a good choice for the combination model of different attention, which can be of the same type \cite{65_pagerank_2021, 19_SuperGAT_2021} or different types \citep{105_RGHAT_2020, 111_HGAT_2021, 140_GMAN_2020}. This combination may make up for the deficiency of a single model, thus improving the expression ability of the model. For example, SuperGAT takes advantage of different attention operations to learn label agreement and edge presence \citep{19_SuperGAT_2021}. Typically, methods with hierarchical attention tend to include two and more attention mechanisms, which can capture different levels of semantic information. The hierarchical attention mechanism utilizes the neighborhood information of an entity more effectively from different levels \citep{105_RGHAT_2020}. For HAN in heterogeneous graphs, the node-level attention aims to learn the importance between a node and its neighbors along the meta-path, while the semantic-level attention can learn the importance of different meta-paths \citep{43_HAN_2019}. More complexly, GMAN \citep{140_GMAN_2020} proposes a graph multi-attention network for long-term traffic prediction, which includes spatial attention, temporal attention, and gated fusion. However, the simultaneous use of multiple attention mechanisms in the same model will also significantly increase the computational complexity. In short, it is difficult to adapt any model to all scenarios. We hope to help researchers better design their attention-based GNNs.

To the best of our knowledge, it is difficult to quantitatively compare and analyze the complexity of different models, especially when different model structures and different attention mechanisms are involved. Therefore, we provide a general analysis of three basic attention-based GNNs. The time complexity of GRANs is high due to their recursive nature. The updating of node representations needs to consider the representations of their neighboring nodes at each time step, and this process requires multiple iterations over time steps. GATs have a high time complexity because they leverage self-attention mechanisms to compute attention weights between nodes. This node-to-node attentional computation increases the space-time complexity. The attention weights can be computed in parallel, which can help to reduce the time complexity but increases the space complexity at the same time. In addition, the scale of graphs continues to grow exponentially over time in real life. This will be a huge challenge for GATs, as well as for GRANs and Graph Transformers. Graph Transformers, based on the Transformer model, involve self-attention and multi-head self-attention in neural network layers. The attention mechanism in Graph Transformers can be seen as a fully connected GAT, and we can imagine that its complexity is usually much higher than that of GAT. The actual time complexity depends on factors such as the specific model architecture, graph size, sparsity, and available computational resources.

\section{Open Issues and Future Directions}
\label{Open Issues and Future Directions}

\subsection{Scalability}
Scalability is a major challenge for the application of attention-based GNNs in practical scenarios. Although there are some open source medium or large-scale graph datasets in OGB \citep{165_Ogb_2020}, most of the existing studies focus on testing the new developed attention-based GNNs on small graphs (e.g., some citation network datasets with only a few thousand nodes such as Cora and Citeseer \citep{2_GCN_2017}), while ignoring the huge, complex and noisy networks in practical applications. 
Large-scale graph data has been a huge challenge for attention-based GNNs. The most fundamental reason for this phenomenon in attention-based GNNs is the high computational complexity of the attention mechanism. Attention-based GNNs are hard to implement in large graphs, especially for Graph Transformers. Up to now, although some approaches have attempted to improve model efficiency through sampling and subgraphs \citep{24_Graph-bert_2020,25_UGformer_2019,159_HGT_2020}, they still can not efficiently process large graphs.

\subsection{Interpretability}
Though attention-based GNNs achieve promising performance on various tasks, a clear understanding of their discriminative power is superficial \citep{2020_GATs_CPA}. Without understanding the relationships behind the predictions, these models can not be understood and fully trusted \citep{167_Explainability_2021}. Recent works \citep{17_how-powerful_2018, 2019_provably} attempt to explore the interpretability of graph neural networks and theoretically analyze the expressive power of GNNs. To improve the performance of attention-based GNNs, CPA \citep{2020_GATs_CPA} improves the attention mechanism in GNNs via cardinality preservation and presents a theoretical analysis of the representational properties of attention-based GNNs. HAGERec proposes a hierarchical attention graph convolutional network to explore users’ potential preferences from the high-order connectivity structure \citep{2020_HAGERec}. Considering the topology of the graph, some researches are carried out from the perspective of motifs \citep{2020_motif} and subgraph \citep{167_Explainability_2021}. However, there is still a lot of room for the interpretability of attention-based GNNs \citep{166_Explain-GNN-survey_2020}.

\subsection{Deeper Models}
Like traditional GNNs, most attention-based GNNs aggregate messages from local neighbors iteratively, ignoring messages from the distant neighborhood. When the downstream task depends on long-range interaction, GNNs fail to propagate messages originating from distant neighborhoods and perform poorly. Although they can obtain distant messages by stacking the neural network layers, this also brings about the over-smoothing \citep{144_Over-smoothing_2018}, i.e., node representations become indistinguishable. To improve expressive capability and alleviate the over-smoothing of attention-based GNNs, some researchers focus on the topological properties of graphs \citep{2020_GATs_CPA, 64_MS-heterophily_2021}. Focus on the architecture of GNNs, decoupling representation transformation and message propagation enable deeper GNNs with attention to learning graph node representations from larger receptive fields \citep{48_DAGNN_2020, 49_GAMLP_2022}. As the number of layers increases, the receptive field of a node grows exponentially. This causes over-squashing \citep{2020_over_squashing}, where information from the exponentially-growing receptive field is compressed into fixed-length node vectors. Compared with GCNs, GATs can effectively alleviate the over-squashing phenomenon by introducing an attention mechanism \citep{2020_over_squashing}. 
However, when stacking too many layers, GATs also suffer from higher complexity. Instead, Graph Transformers \citep{24_Graph-bert_2020, 25_UGformer_2019} obtain global information in a network layer, which is a breakthrough to alleviate the problem of over-squashing. In recent years, many researchers explore the above problems from the perspective of higher-order structures \citep{30_transformers-generalize_2021, 2021_higher_order}, such as hypergraphs and simplicial complexes.

\subsection{Complex Graph}
Graph-structured data is ubiquitous in our daily life, such as social networks, citation networks, and collaboration networks. In the real-world, however, graphs can be both structurally large and complex \citep{20_attention-survey_2019}. In addition to the above graph-structured data often used in scientific research, there are many more complex graph scenarios with a large amount of semantic information used in industry, such as transportation networks \citep{140_GMAN_2020}, knowledge graphs \citep{91_KGAT_2019}, and chemical molecule graphs \citep{37_PAGAT_2019}. These real-world networks are usually modeled as homogeneous graphs \citep{124_heter-survey_2022}, relation-aware graphs \citep{94_RelGNN_2021}, or dynamic graphs \citep{168_ED-survey_2022}. However, in the case of heterogeneous graphs, attention-based GNNs require specific modifications to effectively integrate and combine the semantic information across nodes, edges, and graphs. 
For dynamic graphs with time attributes, the graph structure evolves with time, which increases the difficulty of attention-based GNNs training. Therefore, how to design a more universal attention-based GNN, friendly to graphs with different complex attributes, is still an open problem.

\subsection{Novel Applications}
Attention-based GNNs are widely used in social networks , natural language processing , recommendation systems , and traffic forecasting . In CV that seems to be unrelated to the graph, attention-based GNNs have also begun to be widely applied by generating graphs . 
Attention-based GNNs are widely used in social networks \citep{4_comprehensive-survey_2022}, natural language processing \citep{169_GNN-nlp-survey_2021}, recommendation systems \citep{170_GNN-recommender-survey_2020}, traffic forecasting \citep{171_GNN-traffic_2021}, and multimodal \citep{2023_GNN4Multimodal_survey}. In CV that seems to be unrelated to the graph, attention-based GNNs have also begun to be widely applied by generating graphs \citep{138_T-GNN_2022}. Similarly, we can model segment-wise speaker embeddings (SSEs) as nodes of a graph \citep{2021_GNN4Speaker}. Further, GNNs could model interactions both within and across different data types in multimodal data, including image, video, language, and knowledge graph. Once complex relations between modalities can be built into a network structure, GNNs provide a powerful and flexible strategy to leverage interdependencies in multimodal datasets \citep{2023_GNN4Multimodal_survey}. 
Combinatorial optimization is a canonical NP-hard problem in traditional operations research. GNN as a scalable general-purpose solver is adopted to approximately solve combinatorial optimization problems \citep{172_PIG-GNN_2022}. In the field of biochemistry, GNNs also show unique advantages, especially in generating molecular graphs \citep{173_3D_2021}. It can be seen that attention-based GNNs have good application prospects. There are still many new fields waiting for us to explore, especially in scenes that can be modeled with graphs. 

\section{Conclusion}
\label{Conclusion}

Over the past decade, the attention mechanism has achieved impressive performance in NLP and CV. Attention-based GNNs allow us to adaptively aggregate and update representations from the different neighborhoods with local or global attention, even feature fusion attention. In this survey, we provide a comprehensive review of the most recent research efforts on attention-based GNNs. We first give some basic definitions of attention-based GNNs including graph, graph neural networks, and attention mechanisms. Then, we propose a novel two-level taxonomy for attention-based GNNs from the perspective of development history and architectural perspectives. To be specific, we classify existing attention-based GNNs into three stages and then organize them into several sub-categories in each stage according to the model architecture. For each sub-category, we briefly clarify the main characteristics, detail the attention strategies adopted by the representative models, and discuss their advantages and limitations. Furthermore, we outline some open issues and promising future research directions to advance this field. Finally, we intend to share the relevant latest papers about attention-based GNNs at the open resources. We hope this survey will provide a succinct introduction to attention-based GNNs and shed some light on future developments.

\section*{Appendix}
\appendix
For ease of reading, we summarize the abbreviations of the methods and their corresponding full names in Table 5.
\begin{center}
\label{tab:abbreviation}
\tablefirsthead{%
    \multicolumn{2}{l}{\textbf{Table 5} Abbreviations of the methods and their corresponding full names.}\\
\hline}
\tablehead{%
    \hline
	\multicolumn{2}{l}{\small\sl continued from previous page}\\
	\hline}
\tabletail{%
	\hline
	\multicolumn{2}{r}{\small\sl continued on next page}\\
	\hline}
\tablelasttail{\hline}
\begin{supertabular}{c|l}
Abbreviation & \multicolumn{1}{c}{Full Name}                                                                                                                      \\ \hline
SpectralCNN  & Spectral Convolutional Neural Networks                                                                                                             \\
ChebNet      & Graph CNN with Chebyshev Filter                                                                                                                    \\
GCN          & Graph Convolutional Networks                                                                                                                       \\
MPNN         & Message-Passing Neural Networks                                                                                                                    \\
GraphNet     & Graph Networks                                                                                                                                     \\
GAT          & Graph Attention Networks                                                                                                                           \\
C-GAT        & Constrained Graph Attention Networks                                                                                                               \\
SuperGAT     & Self-supervised Graph Attention Network                                                                                                            \\
GRU          & Gated Recurrent Units                                                                                                                              \\
GGNN         & Gated Graph Sequence Neural Networks                                                                                                               \\
GRNN         & Graph Recurrent Neural Networks                                                                                                                    \\
GSP          & graph signal processing                                                                                                                            \\
GaAN         & Gated Attention Networks                                                                                                                           \\
GRAN         & Graph Recurrent Attention Networks                                                                                                                 \\
GraphSAGE    & Graph SAmple and aggreGatE                                                                                                                         \\
LSTM         & Long Short-Term Memory                                                                                                                             \\
JK-Net       & Jumping Knowledge Networks                                                                                                                         \\
GAM          & Graph Attention Model                                                                                                                              \\
DMP          & Diverse Message Passing                                                                                                                            \\
PPRGAT       & Personalized PageRank GAT                                                                                                                          \\
GANet        & Graph Attention Networks                                                                                                                           \\
CAT          & Conjoint Attention Networks                                                                                                                        \\
MSNA         & MultiSpace Neighbor Attention                                                                                                                      \\
CPA          & Cardinality Preserved Attention                                                                                                                    \\
AGNN         & Attention-based Graph Neural Network                                                                                                               \\
HTNE         & Hawkes process based Temporal Network Embedding                                                                                                    \\
HAT          & Hyperbolic Graph Attention Network                                                                                                                 \\
HGCN         & Hyperbolic Graph Convolutional Neural Network                                                                                                      \\
Hype-HAN     & Hy-perbolic Hierarchical Attention Network                                                                                                         \\
SPAGAN       & Shortest Path Graph Attention Network                                                                                                              \\
PaGNN        & Path-aware Graph Neural Network                                                                                                                    \\
ADSF         & Adaptive Structural Fingerprint                                                                                                                    \\
MAGNA        & Multi-hop Attention Graph Neural Networks                                                                                                          \\
T-GAP        & Temporal Knowledge Graph with Attention Propagation                                                                                                \\
SiGAT        & Signed Graph Attention Networks                                                                                                                    \\
SNEA         & Signed Network Embedding via Graph Attention                                                                                                       \\
RGAT         & Relational Graph Attention Networks                                                                                                                \\
HetSANN      & Heterogeneous Graph Structural Attention Neural Network                                                                                            \\
WRGAT        & Weighted Relational GAT                                                                                                                            \\
EAGCN        & Edge Attention-based Multi-Relational GCN                                                                                                          \\
TALP         & Type-Aware Anchor Link Prediction                                                                                                                  \\
KGAT         & Knowledge Graph Attention Network                                                                                                                  \\
GATE         & Graph Attention Transformer Encoder                                                                                                                \\
RelGNN       & Relation-aware GNN                                                                                                                                 \\
CGAT         & Channel-aware Graph Attention Network                                                                                                              \\
PSHGAN       & \begin{tabular}[c]{@{}l@{}}Meta-path and meta-structure integrated heterogeneous\\  graph neural network through attention mechanisms\end{tabular} \\
PRML         & \begin{tabular}[c]{@{}l@{}}Path-based Proximity Ranking Metric \\ Dual-Level Attention Learning\end{tabular}                                       \\
RGHAT        & \begin{tabular}[c]{@{}l@{}}Relational Graph neural network \\ with Hierarchical ATtention\end{tabular}                                             \\
LAN          & Logic Attention Network                                                                                                                            \\
GraphHAM     & \begin{tabular}[c]{@{}l@{}}Graph embedding model \\ with Hierarchical Attentive Memberships\end{tabular}                                           \\
GAW          & Graph Attention model with random Walk                                                                                                             \\
NLGAT        & Non-Local GAT                                                                                                                                      \\
DiffPool     & Differentiable Graph Pooling                                                                                                                       \\
SAGPool      & Self-Attention Graph Pooling                                                                                                                       \\
AttPool      & \begin{tabular}[c]{@{}l@{}}graph pooling module based on \\ an attention based mechanism\end{tabular}                                              \\
ChebyGIN     & Combining GIN with ChebyNet                                                                                                                        \\
DAGNN        & Deep Adaptive Graph Neural Network                                                                                                                 \\
TDGNN        & Tree Decomposed Graph Neural Network                                                                                                               \\
FAGCN        & Frequency Adaptation Graph Convolutional Networks                                                                                                  \\
ACM          & Adaptive Channel Mixing                                                                                                                            \\
AM-GCN       & Adaptive Multi-channel GCN                                                                                                                         \\
UAG          & Uncertainty-aware Attention Graph Neural Network                                                                                                  \\
MV-GCN       & Multi-View Graph Convolutional Netowork                                                                                                            \\
GENet        & Graph Ensemble Network                                                                                                                             \\
MVE          & Multi-View Network Representation Learning                                                                                                         \\
MGAT         & Multi-view Graph Attention Networks                                                                                                                \\
DySAT        & Dynamic Self-Attention Network                                                                                                                     \\
TemporalGAT  & Temporal-GAT                                                                                                                                       \\
GAEN         & Graph Attention Evolving Networks                                                                                                                  \\
MMDNE        & Network Embedding with Micro- and Macro-Dynamics                                                                                                   \\
TGAT         & Temporal Graph Attention                                                                                                                           \\
MTAD-GAT     & \begin{tabular}[c]{@{}l@{}}Multivariate Time-series Anomaly Detection \\ via Graph Attention Network\end{tabular}                                  \\
GACNN        & Graph Attention Convolutional Neural Network                                                                                                       \\
Hyper-SAGNN  & \begin{tabular}[c]{@{}l@{}}Self-Attention based Graph Neural Network \\ for Hypergraphs\end{tabular}                                               \\
HHGR         & \begin{tabular}[c]{@{}l@{}}Hierarchical Hypergraph Learning Framework \\ for Group Recommendation\end{tabular}                                     \\
HyperTeNet   & Hypergraph and Transformer-based Neural Network                                                                                                    \\
Hyper-GAT    & Hypergraph-GAT                                                                                                                                     \\
PAGAT        & Path-Augmented Graph Transformer Networks                                                                                                          \\
GT           & Graph Transformer                                                                                                                                  \\
SAN          & Spectral Attention Network                                                                                                                         \\
GTA          & Graph Truncated Attention                                                                                                                          \\
UniMP        & Unified Message Passaging Model                                                                                                                    \\
HOT          & Higher-Order Transformer                                                                                                                           \\
GraphFormers & GNN-nested Transformers                                                                                                                            \\
UGformer     & Universal Graph Transformer                                                                                                                        \\
HGT          & Heterogeneous Graph Transformer                                                                                                                    \\
GROVER       & \begin{tabular}[c]{@{}l@{}}Graph Representation frOm self-superVised \\ mEssage passing tRansformer\end{tabular}                                   \\
GTN          & Graph Transformer Networks                                                                                                                         \\
GMT          & Graph Multi-set Transformer                                                                                                                        \\
OGB          & Open Graph Benchmark                                                                                                                               \\ \hline
\end{supertabular}
\end{center}

\section*{Acknowledgements}
This work was supported by the National Natural Science Foundation of China (No.61876186) and the Xuzhou Science and Technology Project (No.KC21300).


\bibliography{Manuscript}


\end{document}